\documentclass[pdflatex,sn-mathphys-num]{sn-jnl}

\usepackage{graphicx}%
\usepackage{multirow}%
\usepackage{amsmath,amssymb,amsfonts}%
\usepackage{amsthm}%
\usepackage{mathrsfs}%
\usepackage[title]{appendix}%
\usepackage{xcolor}%
\usepackage{textcomp}%
\usepackage{manyfoot}%
\DeclareNewFootnote{A}[gobble]%
\usepackage{booktabs}%
\usepackage{algorithm}%
\usepackage{algorithmicx}%
\usepackage{algpseudocode}%
\usepackage{listings}%

\usepackage{natbib}
\usepackage{dsfont}
\usepackage{subcaption}
\usepackage{color}
\usepackage{enumerate}
\usepackage{cases}
\usepackage{multicol}
\usepackage{enumitem}
\usepackage{bm}
\usepackage{import}
\usepackage{lscape}
\usepackage[normalem]{ulem}
\usepackage{url}
\usepackage{geometry}
\usepackage{adjustbox}

\definecolor{BrightBlue}{RGB}{65, 145, 225}

\makeatletter
\newcommand\footnoteref[1]{\protected@xdef\@thefnmark{\ref{#1}}\@footnotemark}
\makeatother

\theoremstyle{thmstyleone}%
\newtheorem{theorem}{Theorem}
\newtheorem{proposition}[theorem]{Proposition}%

\theoremstyle{thmstyletwo}%

\theoremstyle{thmstylethree}%
\begin{document}

\title{AI-driven Prices for Externalities and Sustainability in Production Markets}



\author*[1]{\fnm{Panayiotis} \sur{Danassis}}

\author[2,3]{\fnm{Aris} \sur{Filos-Ratsikas}}

\author[4]{\fnm{Haipeng} \sur{Chen}}

\author[5]{\fnm{Milind} \sur{Tambe}}

\author[6]{\fnm{Boi} \sur{Faltings}}

\affil[1]{\orgname{University of Southampton}, \city{Southampton}, \country{UK}}
\affil[2]{\orgname{University of Edinburgh}, \city{Edinburgh}, \country{UK}}
\affil[3]{\orgname{Imperial College London}, \city{London}, \country{UK}}
\affil[4]{\orgname{William \& Mary}, \city{Williamsburg}, \state{Virginia}, \country{USA}}
\affil[5]{\orgname{Harvard University}, \city{Cambridge}, \state{Massachusetts}, \country{USA}}
\affil[6]{\orgname{\'Ecole Polytechnique F\'ed\'erale de Lausanne (EPFL)}, \city{Lausanne}, \country{Switzerland}}



\abstract{Traditional competitive markets do not account for negative externalities; indirect costs that some participants impose on others, such as the cost of over-appropriating a common-pool resource (which diminishes future stock, and thus harvest, for everyone). Quantifying appropriate interventions to market prices has proven to be quite challenging. We propose a practical approach to computing market prices and allocations via a deep reinforcement learning policymaker agent, operating in an environment of other learning agents. Our policymaker allows us to \emph{tune} the prices with regard to diverse objectives such as sustainability and resource wastefulness, fairness, buyers' and sellers' welfare, etc. As a highlight of our findings, our policymaker is significantly more successful in maintaining resource \emph{sustainability}, compared to the market equilibrium outcome, in scarce resource environments.}


\artnote{This article is an extended version of~\cite{DanassisPolicymaker}.}



\maketitle

\section{Introduction} \label{sec:Introduction}

Competitive markets, founded in the works of~\citeauthor{walras1874elements} and \citeauthor{Fisher1892PhDThesis}, constitute the fundamental mechanism of allocation; the means by which products are sold and bought. Market theory \cite{arrow1954existence} suggests that these markets will reach an efficient, stable outcome, the \emph{market equilibrium}, in which supply equals demand, and all participants are maximally satisfied by the bundles of goods that they buy or sell at the chosen prices.

Nevertheless, these established market models fail to account for negative \emph{externalities}~\cite{laffont2008externalities}, which lead to market failure~\cite{bator1958anatomy}. These externalities refer to indirect costs that are not reflected in the market equilibrium prices. A representative example of such inefficiencies is the environmental harm caused by pollution and overexploitation of natural resources, e.g., air pollution from burning fossil fuels, water pollution from industrial effluents, antibiotic resistance due to the overuse of antibiotics in industrial farms, etc. Another prominent example of a negative externality -- which we will use as a real-world, indicative test-case throughout the paper -- is the depletion of the stock of fish due to overfishing.\footnote{For example, according to OECD, about 25\% of fish stocks globally are at risk~\cite{oecd_review_of_fisheries}.} With these ``exogenous'' objectives being of paramount importance, it is only natural to assume some form of intervention to the reign of these markets.

There are many approaches to reconciling the economics of the competitive market with societal and environmental externalities. For example, policy-makers can correct for the inefficiencies by employing command-and-control legislation (e.g.,~\cite{wiki_clean_air_act}), permit markets~\cite{crocker1966structuring,dales1968pollution}  (e.g.,~\cite{eu_ets}), or taxation (e.g.,~\cite{eu_green_taxation}). A classic example of the latter is Pigouvian taxes~\cite{pigou1924economics}, i.e., taxes that are equal to the external damage caused by production decisions. While such interventions are clearly necessary, \emph{selecting and quantifying} the appropriate ones has proven to be quite challenging. For instance, in the case of common fisheries, approaches aiming to determine the ``optimal'' level of annual harvest and subsequently control fishing to achieve that quota have often failed to prevent overfishing~\cite{clark2006worldwide}. Similarly, determining the marginal social cost of a negative externality and converting it into a monetary value can be quite impractical~\cite{baumol1972taxation}. Another approach, founded in the seminal works of Samuelson~\cite{samuelson1954pure,samuelson1995diagrammatic}, attempts to capture and address externalities via the introduction of \emph{public goods} in the market, and the computation of the equilibrium points for these ``extended'' markets. The extent to which these public goods can capture all appropriate government functions has been largely debated (e.g., see~\cite{samuelson1995diagrammatic,pickhardt2006fifty}) and besides, the impracticality of computing the equilibrium points of these markets analytically is even more pronounced than in ``traditional'' competitive markets.  

An added complication when it comes to devising effective policies for sustainability and combating externalities, or any societal objective for that matter, is the fact that the interactions between the different entities in the market ecosystem are rather complex and of a repeated nature. Indeed, the appropriate mathematical modeling of these systems is that of a \emph{stochastic} (or \emph{Markov}) game (see Section~\ref{Models}), in which the actors (i.e., the policymaker and the harvesters of natural resources) are both aiming to optimize their individual utilities over a fixed horizon. To do that, they need to optimize over their future rewards, taking into account the effect of the actions of the other actors on their own. Solving these games analytically is both conceptually and computationally hard, even for relatively simple variants of those games and well-behaved equilibrium notions (e.g., see~\cite{chatterjee2004nash,daskalakis2022complexity,deng2021complexity}). 
For this reason, most classic works in economics and mathematics (e.g.,~\cite{sobel1971noncooperative,himmelberg1976existence,solan2002correlated}) have only gone as far as identifying conditions that merely establish the existence of some equilibrium, without providing any guarantees about its properties. Besides the computational burden, another significant hurdle to the analytical approach is that it typically requires full observability of the environment and the actions of the other participants, which is most often not the case in practice. 

Reinforcement learning (see~\cite{kaelbling1996reinforcement}) has been proposed and extensively used as an alternative approach for computing optimal strategies in stochastic games \cite{Littman2002}. The idea is that the actors, as learners, interact with their environment exclusively via signals of limited information: they typically observe their rewards based on their past actions, and update their current actions accordingly, via the employment of some carefully devised learning algorithm. Note that this approach does not require observable information about the parameters of the environment or the other actors. It also does not require the derivation of analytical solutions to complex optimization problems, as ``off-the-shelf'' reinforcement learning algorithms are readily available. For these reasons, an established line of work has considered reinforcement learning as a form of bounded rationality \cite{rubinstein1998modeling} which is much more conceivable for complex environments in practice, compared to the standard ``perfect'' rationality of traditional economic agents (see Section~\ref{sec:relatedworkLearningAgents} for more details). Finally, reinforcement learning has been shown to be generally robust to changes in a range of input parameters, making it very suitable for complex and volatile environments.

Motivated by the arguments above, we propose a practical and effective technique for calculating concrete market prices and allocations via a \emph{deep reinforcement learning policymaker}, operating in an environment of other learning agents. These prices can serve as a clear-cut guideline for intervention, and can then be implemented by a variety of mechanisms; e.g., policy-makers can tax (or subsidize) the difference between the current market price and the computed price, or buy/sell from reserves.\footnote{There are many such examples of influencing the supply/demand e.g., see~\cite{cnn_mapple,wpr_milk,bbc_oil}.} This new approach grants us the ability to abstract real-world situations into a form that makes them amenable to research, and allows for advances in the state-of-the-art. In particular, it enables us to design and test novel policies (via tuning the parameters and simulating the multi-agent environment) to tackle a plethora of real-world problems in various disciplines under a host of objectives, such as the problem of sustainable production (renewable energy, CO\textsubscript{2} markets, natural resource preservation, etc.). Our work falls into the very recent research agenda of building agent-based models to inform policy in socio-economic environments (e.g., see~\cite{zheng2020ai,deepmindEcon}).

\subsection{Our Contributions} \label{Our Contributions}

In this paper, we use deep reinforcement learning for policy making. We study the emergent behaviors as a group of deep learners interact in a complex and realistic market, where both the pricing policy and the harvesting behaviors are learned \emph{simultaneously}. Neither the policymaker nor the harvesters have prior knowledge / assumptions of domain dynamics or economic theory, and every agent only makes use of information that it can individually observe. In particular:

\medskip
\noindent
\textbf{(1) We propose a practical approach to computing market prices and allocations via a deep reinforcement learning policymaker agent} that allows us to \emph{tune} the prices with regard to diverse objectives such as \emph{sustainability} and resource \emph{wastefulness}, \emph{fairness} and buyers' and sellers' welfare. Our goal is to investigate the feasibility of using traditional deep reinforcement learning agents as (i) a practical alternative to classical notions of rationality and market equilibria, and (ii) a means to reach stable outcomes that are comparable with the idealized market equilibrium outcome from economics, while at the same time optimizing exogenous objectives.

\medskip
\noindent
\textbf{(2) We introduce a novel multi-agent socio-economic environment and \emph{prove} a sufficient condition for market failure}. Our environment combines established principles of competitive markets with the challenges of \emph{resource scarcity} and the tragedy of the commons. This is paramount to understanding the impact of self-interested appropriation and developing sustainable strategies. While we use a common fishery as an indicative, real-world test-bench, our approach is general and can be employed in \emph{any} production market. To demonstrate the need for intervention via our policymaker, we provide an analytical example (see Section \ref{sec:exampleMarketFailure}) where leaving the market entirely ``free'' to act according to the market equilibrium will result in the depletion of the resource in a short period of time, even under optimal harvesting. Most importantly, our analysis also highlights the inherent challenges of theoretically computing optimal strategies for either the policymaker or the harvesters (\textbf{even under strong assumptions, computing the optimal policy analytically is impractical}), and justifies modeling both as learning agents instead.

\medskip
\noindent
\textbf{(3) We provide a thorough (quantitative \& qualitative) analysis} on the learned policies and demonstrate that they can achieve significant improvements over the market equilibrium benchmark for several objectives, while maintaining comparable performance for the rest. As a highlight of our results, we show that our policymaker fares notably better in terms of the sustainability of resources, essentially without compromising any of the remaining objectives.

Given that it is often quite hard to experiment with real-world pricing policies, traditional work in economics often results in simplifying assumptions which are hard to validate. Our approach provides an alternative route, enabling experimentation (via tuning of the parameters and simulating the multi-agent environment) to find the best possible policies.

\subsection{Discussion \& Related Work} \label{Related Work}

\subsubsection{Competitive Markets} 

The origins of competitive market theory date back to the pioneering ideas of~\citeauthor{walras1874elements} and~\citeauthor{Fisher1892PhDThesis}. \citeauthor{arrow1954existence}, defined and studied the standard, most general model of competitive markets, and proved the existence of a market equilibrium (ME). The market that we consider in this paper is a special case of the Arrow-Debreu model, due to Fisher \cite{Fisher1892PhDThesis}, where the market participants are divided into buyers and sellers, and buyers do not have intrinsic value for money, but rather use money as a means of facilitating the trade. We chose the (linear) ``Fisher market'' as our benchmark because, contrary to general Arrow-Debreu markets (e.g., see~\cite{chen2017complexity}), computing a ME can be done in polynomial time. We remark that in our setting the harvesters are the recurring entities in the system, as they return to the market with (potentially) new harvest in each round. On the other hand, we assume that buyers are short-lived (i.e., they only participate in the market once) or myopic (i.e., they are trying to maximize their utility at the current market stage). \citeauthor{goktas2022zero} recently defined the notion of a \emph{stochastic} Fisher market (extending the ideas of \cite{friesen1979arrow} for Arrow-Debreu-type markets), where the same buyers participate in the same Fisher market over a sequence of steps, aiming to maximize their long-term utility and possibly saving part of their budget for future steps. As our focus in this work is on the harvesters (who are also our learning agents) rather than the buyers, we elected to study the standard, static Fisher market model instead.

The ME is theoretically reached via the continuous adjustment of supply and demand dictated by the market's ``invisible hand''~\cite{smith1791inquiry} (``t\^{a}tonnment process''). Yet, note that the ME is reached only under a strict range of assumptions (e.g., participants are price-takers, there is no collusion, etc.), and the t\^{a}tonnment process is highly dependent on several initial parameters and can therefore be slow, and even impractical to compute~\cite{codenotti2006leontief,chen2017complexity,chen2009settling}. We also remark that while similar in spirit, the ME is a different notion from the well-known notion of the Nash equilibrium~\cite{nash1950equilibrium}; the former is a stable point of the market supply and demand adjustment, whereas the latter is a stable point of the participants' strategic play. In particular, the classic ME results assume that agents are \emph{not strategic} and therefore do not attempt to influence the prices of the markets (price-takers). In the presence of rational agents, the outcome of the market can be fundamentally different from the ME \cite{branzei2014fisher,adsul2010nash,chen2016incentives}.

\subsubsection{Learning Agents}\label{sec:relatedworkLearningAgents}

The fundamental assumptions of price-taking and perfect rationality have been challenged in recent years by the emergence of \emph{learning agents}. As autonomous agents proliferate, they will be called upon to interact in ever more complex environments, and as such, will play a key part in sustainable production. In fact, learning agents have become ubiquitous in socio-economic and socio-ecological systems in recent years (e.g.,~\cite{10.5555/3463952.3464003,zheng2020ai}). This has led to the emergence of machina economica~\cite{Parkes267}, an approximate counterpart of homo economicus -- the perfectly rational
agent of neoclassical economics -- given computational barriers and the lack of common knowledge. For example, with the emergence of machine learning, it has been observed (e.g., see~\cite{tardos2019learning}) that enterprises use learning agents as forms of bounded rationality \cite{rubinstein1998modeling}. These can range from simple no-regret bandit algorithms like in \cite{cai2018reinforcement,nekipelov2015econometrics}, to more complex approaches. The success of multi-agent deep reinforcement learning (DRL) has led to a growing interest in modeling machinae economicae agents as \emph{independent} deep reinforcement learning agents, learning from observational data alone without any economic modeling assumptions. As an example, in a recent work, \citeauthor{zheng2020ai} use DRL to study taxation policies, but in a simpler domain (they use a synthetic appropriation game -- compared to our real-world fishery model -- and do not consider a concrete market model like our Fisher market). Moreover, the resources can not be permanently depleted, thus they do not address the challenge of sustainability and the tragedy of the commons, and they allow for weight sharing during training, while we have completely independent learning agents. There is other recent work that has adopted a similar agenda, but in markedly different domains and using different approaches~\cite{pmlr-v97-duetting19a,shen2019automated,cai2018reinforcement,deepmindEcon,vadori2022towards}. Our work is one of the first to design a \emph{practical} policymaker via deep reinforcement learning in realistic economic environments.

Finally, there has been great interest lately in Common-Pool Resource (CPR) appropriation problems as an application domain for Multi-agent Deep Reinforcement Learning (MADRL), e.g.,~\cite{zheng2020ai,NEURIPS2020_ad7ed5d4,DanassisJaamasFisheries}. CPR problems offer complex environment dynamics and relate to real-world socio-ecological systems. \citeauthor{10.5555/3463952.3464003} were the first to introduce the realistic fishery model that we employ in this work (see also,~\cite{DanassisJaamasFisheries,danassis2022Scalable}). We extended the model to deal with multiple resources, harvesters with diverse skill levels, and a realistic Fisher market.

\section{Environment and Agent Models} \label{Models}

In this section, we provide a detailed description of our complex economic model. It consists of (i) a common-pool resource appropriation game -- where a group of appropriators compete over the harvesting of a set of common resources and which exhibits properties related to the tragedy of the commons \cite{hardin1968tragedy} and the challenge of \emph{sustainability} (resource scarcity) -- and (ii) a complex and realistic market (with a dynamic set of buyers and sellers, endowments, and utilities), where the appropriators sell their harvest.

\subsection{The Common Fishery Model} \label{Fishery Model}

Fisheries constitute a common-pool resource of finite yield (i.e., it is challenging and/or costly to exclude individuals from appropriating), thus they are vulnerable to the \emph{tragedy of the commons} \cite{hardin1968tragedy}. Fishermen do not consider the costs to others; they harvest more than is efficient (i.e., they deplete the resource faster than it can regenerate), leading to environmental degradation (ecological market failure) and may eventually threaten entire ecosystems with extinction (e.g., see the Atlantic cod fishery~\cite{daly2011ecological}).

We adopt the fishery model of \cite{10.5555/3463952.3464003}, which is based on an abstracted bio-economic model for \emph{real-world} commercial fisheries \cite{clark2006worldwide,diekert2012tragedy}. We chose this environment due to its \emph{complex dynamics}, but the proposed approach can be employed in \emph{any market and for any resources}, not just fisheries. We have extended the model to account for \emph{multiple resources} and harvesters with \emph{varying skill levels}. Our model describes the dynamics of the stock of a set of common-pool resources, as a group of appropriators harvest over time. The harvest depends on (i) the effort exerted, and (ii) the ease of harvesting that particular resource at that point in time, which depends on its stock level. The stock replenishes at a rate dependent on the current stock level.

More formally, let $\mathcal{N}$ denote the set of appropriators (harvesters) and $\mathcal{R}$ the set of resources. Let $\bm{\eta}_{n} = [\eta_{n, 1}, \dots, \eta_{n, r}, \dots, \eta_{n, R}]$, where $\eta_{n, r} \in [0, 1]$ denotes the skill\footnote{In our model $\eta_{n, r}$ does not depend on time, but one can consider agents that increase their skill level as they harvest. Moreover one can introduce heterogeneous social groups and consider the problem of social mobility.} (competence) of harvester $n$ for harvesting resource $r$. At each time-step $t$, every agent exerts a vector of efforts $\bm{\phi}_{n, t} = [\phi_{n, 1, t}, \dots, \phi_{n, r, t}, \dots, \phi_{n, R, t}]$, where $\phi_{n, r, t} \in [0, \Phi_{\max}]$ is the effort exerted to harvest resource $r$.

Let $\bm{\varepsilon}_{n, t} = \bm{\phi}_{n, t} \cdot \bm{\eta}_{n} = [\varepsilon_{n, 1, t}, \dots, \varepsilon_{n, r, t}, \dots, \varepsilon_{n, R, t}]$ denote the ``effective effort'', and $E_{r, t} = \sum_{n \in \mathcal{N}} \varepsilon_{n, r, t}$ the total effort exerted by all the harvesters at resource $r$ at time-step $t$. Then, the total harvest of resource $r$ is given by Equation \ref{Eq: Total harvest}, where $s_{r, t} \in [0,\infty)$ denotes the stock level at time-step $t$, $q_r(\cdot)$ denotes the catchability coefficient (Equation \ref{Eq: Catchability coefficient function}), and $S_r^{eq}$ is the equilibrium stock of the resource.
\begin{equation} \label{Eq: Total harvest}
  H_r(E_{r, t}, s_{r, t}) =
  \begin{cases}
    q_r(s_{r, t}) E_{r, t} &, \text{if } q_r(s_{r, t}) E_{r, t} \leq s_{r, t} \\
    s_{r, t} &, \text{otherwise} \\
  \end{cases}
\end{equation}
\begin{equation} \label{Eq: Catchability coefficient function}
  q_r(x) = 
  \begin{cases}
    \frac{x}{2 S_r^{eq}} &, \text{if } x \leq 2 S_r^{eq} \\
    1  &, \text{otherwise} \\
  \end{cases}
\end{equation}

Each environment can only sustain a finite amount of stock. If left unharvested, the stock will stabilize at the equilibrium stock $S_r^{eq}$. Note also that $q_r(\cdot)$, and therefore $H_r(\cdot)$, are proportional to the current stock, i.e., the higher the stock, the larger the harvest for the same total effort. The stock dynamics of each resource are governed by Equation \ref{Eq: stock dynamics}, where $F(\cdot)$ is the spawner-recruit function (Equation \ref{Eq: Spawner-recruit function}) which governs the natural growth of the resource, and $g_r$ is the growth rate.\footnote{Assuming $s_0 \leq 2 S^{eq}$, and following the derivation of \cite{10.5555/3463952.3464003}, to avoid highly skewed growth models and unstable environments we need $g_r \leq -W_{-1}(-1/(2 e)) \approx 2.678$, where $W_k(\cdot)$ is the Lambert $W$ function.}
\begin{equation} \label{Eq: stock dynamics}
  s_{r, t+1} = F(s_{r, t} - H_r(E_{r, t},s_{r, t}))
\end{equation}
\begin{equation} \label{Eq: Spawner-recruit function}
    F(x) = x e^{g_r (1 - \frac{x}{S_r^{eq}})}
\end{equation}

We assume that the individual harvest is proportional to the exerted effective effort\footnote{When no effective effort is exerted, i.e., $E_{r,t}=0$, the individual harvest is defined to be zero.} (Equation \ref{Eq: Individual harvest}), and the revenue of each appropriator is given by Equation \ref{Eq: Revenue}, where $p_{r, t} \geq 0$ is the price (\$ per unit of resource), and $c_{n, r, t}(\cdot)$ is the cost (\$) of harvesting (e.g., operational cost, taxes, etc.). Here lies the ``tragedy'': the benefits from harvesting are private ($p_{r, t} h_{n, r, t}(\cdot)$), but the loss is borne by all (in terms of a reduced stock, see Equation \ref{Eq: stock dynamics}).

\begin{equation} \label{Eq: Individual harvest}
    h_{n, r, t}(\varepsilon_{n, r, t}, s_{r, t}) =
    \begin{cases}
        \frac{\varepsilon_{n, r, t}}{E_{r, t}} H_r(E_{r, t}, s_{r, t}) & \text{if } E_{r,t} > 0,\\
        0, & \text{if } E_{r,t} = 0.
    \end{cases}
\end{equation}

\begin{equation} \label{Eq: Revenue}
  u_{n, r, t}(\varepsilon_{n, r, t}, s_{r, t}) = p_{r, t} h_{n, r, t}(\varepsilon_{n, r, t}, s_{r, t}) - c_{n, r, t}(\varepsilon_{n, r, t})
\end{equation}

Having a realistic environment that exhibits resource scarcity, ``the tragedy of the commons'', and the challenge of sustainability is not only important for the sake of realism, but it can potentially drastically impede the learning process. The benefits of harvesting can lead to greedy agents, which in turn deplete the resources \emph{early} in the episode. This will result in short episodes, limiting the learning per episode.\footnote{In contrast to alternative CPR games in the literature (e.g.,~\cite{zheng2020ai}) where resources respawn after being depleted.}

\subsection{The Fisher Market Model} \label{Market Model}

In a Fisher market there is a set of buyers $\mathcal{B}$ and a set of divisible goods (resources) $\mathcal{R}$, sold by one or multiple sellers. Every seller brings to the market a quantity of each good, with $e_r$ denoting the total quantity of good $r \in \mathcal{R}$ brought collectively by the sellers. Every buyer brings a monetary endowment, or simply a \emph{budget} of $\beta_b$, for $b \in \mathcal{B}$. Additionally, every buyer $b$ has a \emph{valuation} $v_{b, r}$ for each unit of good $r$. An allocation $\mathbf{x}$ is an $|\mathcal{B}| \times |\mathcal{R}|$ matrix, where $x_{b, r}$ denotes the amount of good $r$ that is allocated to buyer $b$. In a feasible allocation, it holds that $\sum_{b \in \mathcal{B}} x_{b, r} \leq e_r$, for any good $r$. We will consider linear Fisher markets, where the \emph{utility} of a buyer given allocation $\mathbf{x}$ is defined as:
\begin{equation} \label{Eq: market buyers utility}
  u_b(\mathbf{x}) = \sum_{r \in \mathcal{R}} x_{b, r} v_{b, r}
\end{equation}

\subsubsection{Market Equilibrium} \label{sec:Market Equilibrium}

These markets are also often called \emph{Eisenberg-Gale Markets} \cite{eisenberg1959consensus}.\footnote{Strictly speaking, the term ``Eisenberg-Gale Market'' is often used to refer to Fisher markets with CES utility functions, which are a superclass of the linear utility functions that we consider here.} In such markets, a \emph{(competitive) market equilibrium} is a pair $(\mathbf{x}, \bm{p})$ of an allocation and a vector of prices, one for each good, such that at these prices each buyer is allocated a utility-maximizing bundle of goods, the budgets are entirely spent, and the goods are entirely sold (market clearance). For Eisenberg-Gale markets in particular, a market equilibrium can be found as the solution to the following convex optimization program:
\begin{eqnarray} \label{Eq: convex optimization program}
  \max \ \ & & \sum_{b \in \mathcal{B}} \beta_b \cdot \log(u_b) \\
  s.t. \ \ & & u_b=\sum_{r \in \mathcal{R}} v_{b, r}\cdot x_{b, r},\ \ \forall \ b \in \mathcal{B} \nonumber \\
  & & \sum_{b \in \mathcal{B}} x_{b, r}\leq e_r,\ \ \forall \ r \in \mathcal{R} \nonumber \\
  & & x_{b, r}\geq 0,\ \ \forall \ b \in \mathcal{B}, \ \ r \in \mathcal{R}
  \nonumber
\end{eqnarray}

While the prices do not strictly appear in this formulation, they can be recovered as the Lagrangian multipliers for the second set of constraints (the feasibility constraints of the good supply). Given the above formulation, a market equilibrium in a Fisher market always exists and it can also be computed in polynomial time. In fact, there are also combinatorial algorithms for equilibrium computation in Fisher markets, e.g., see \cite{jain2010eisenberg,bei2016computing,chakrabarty2006new}.

\subsection{Market Failure Under Optimal Harvesting} \label{sec:exampleMarketFailure}

In this section we provide concrete examples of conditions that lead to market failure -- specifically, the depletion of the resource -- under market equilibrium prices and \emph{optimal harvesting}. We remark that while market failures are a well-documented phenomenon \cite{randall1983problem,sweeney1974market,laffont2008externalities,baumol1972taxation}, our example shows that they can actually happen even when the harvesters are coordinating and harvesting using \emph{optimal strategies}, which in fact incorporate the need for future harvest, to ensure not just immediate but also future rewards.

To simplify our analysis, we consider a setting where a single resource is harvested and sold and which the means of harvesting are controlled by a single entity;  effectively, the latter corresponds to having a single harvester in our model. Thus, we only have a single control variable, $\varepsilon_t$, corresponding to the cumulative effort/harvest.\footnote{To improve readability, we have omitted the subscripts $n$, and $r$, in this section.} This makes our example stronger, as we show that even if the harvesters could somehow conceivably coordinate to harvest optimally as a group, the resource would still be depleted.

\bigskip
\begin{theorem}[Bellman characterization of the optimal harvesting policy]
\label{thm:bellman_characterization}
Consider the single-resource, single-controller problem over a finite horizon $T$, with state $s_t \in [0,\infty)$, control $\varepsilon_t \in [0,E_{\max}]$, stock dynamics
\[
s_{t+1}=F\bigl(s_t-H(\varepsilon_t,s_t)\bigr),
\]
and one-period payoff (according to Equation \ref{Eq: Revenue})
\begin{equation} \label{Eq: reward}
    u_t(s_t,\varepsilon_t)=p_t H(\varepsilon_t,s_t)-c_t(\varepsilon_t),
\end{equation}
where $p_t \ge 0$ is exogenous and $c_t(\cdot)$ is continuously differentiable. 
Then the value function $V_t$ satisfies the Bellman recursion (where $\gamma \in (0,1]$ denotes the discount factor)
\[
V_t(s)=\max_{\varepsilon \in [0,E_{\max}]}
\left\{
p_t H(\varepsilon,s)-c_t(\varepsilon)+\gamma V_{t+1}\bigl(F(s-H(\varepsilon,s))\bigr)
\right\}.
\]
Assume additionally that, at time $t$, the system lies in the uncapped-harvest region, i.e., 
\[
H(\varepsilon,s)=q(s) \varepsilon,
\]
with $q(s)>0$. Then any interior maximizer $\varepsilon_t^*(s)\in(0,E_{\max})$ must satisfy
\begin{equation}
\label{eq:bellman_foc}
p_t q(s)-c_t'\!\bigl(\varepsilon_t^*(s)\bigr)
=
\gamma q(s)\,
V_{t+1}'\!\left(F\bigl(s-q(s)\varepsilon_t^*(s)\bigr)\right)
F'\!\left(s-q(s)\varepsilon_t^*(s)\right).
\end{equation}
Equivalently, defining the post-harvest stock
\[
w_t^*(s) \triangleq s-q(s)\varepsilon_t^*(s),
\]
any interior optimum satisfies
\begin{equation}
\label{eq:bellman_foc_w}
\gamma
V_{t+1}'\!\left(F(w_t^*(s))\right)F'(w_t^*(s))
=
p_t-\frac{c_t'\!\bigl(\varepsilon_t^*(s)\bigr)}{q(s)}.
\end{equation}
\end{theorem}

In general, the optimal policy does not admit a closed-form expression, because the first-order condition depends on the derivative of the value function $V'_{t+1}$.

\begin{proof}
To improve readability, we have omitted the subscripts $n$, and $r$, since we consider a setting where a single resource is harvested and sold and which the means of harvesting are controlled by a single entity.

The Bellman recursion is standard for finite-horizon dynamic programming. Let $V_t(s)$ denote the value function, i.e., the maximal discounted payoff from time $t$ onward when the current stock is $s$. Since the problem has finite horizon $T$ and no terminal payoff, the value function after the last timestep is zero, i.e., $V_{T+1}(s)=0$ for all $s$. The value function is defined as (where $\gamma \in (0,1]$ denotes the discount factor):
\[
V_t(s)=\max_{\varepsilon \in [0,E_{\max}]}
\left\{
u_t(s,\varepsilon)+\gamma V_{t+1}(s')
\right\},
\]
where
\[
s'=F\bigl(s-H(\varepsilon,s)\bigr).
\]
Substituting the definition of $u_t$ (Equation \ref{Eq: reward}) yields
\[
V_t(s)=\max_{\varepsilon \in [0,E_{\max}]}
\left\{
p_t H(\varepsilon,s)-c_t(\varepsilon)+\gamma V_{t+1}\bigl(F(s-H(\varepsilon,s))\bigr)
\right\},
\]
which provides the Bellman equation.

Now suppose that, for the state $s$ under consideration, the system lies in the uncapped-harvest region\footnote{See also \cite{10.5555/3463952.3464003}}, so that
\[
H(\varepsilon,s)=q(s) \varepsilon.
\]
Since $s$ is fixed when optimizing at time $t$, the Bellman objective becomes
\[
\Phi_t(\varepsilon;s) \triangleq p_t q(s) \varepsilon - c_t(\varepsilon) + \gamma V_{t+1}\bigl(F(s-q(s)\varepsilon)\bigr),
\qquad \varepsilon\in[0,E_{\max}].
\]
Let $\varepsilon_t^*(s)\in(0,E_{\max})$ be an interior maximizer. Since $\Phi_t(\cdot;s)$ is differentiable, the first-order necessary condition is
\[
\frac{\partial \Phi_t}{\partial \varepsilon}(\varepsilon_t^*(s);s)=0.
\]
Differentiating $\Phi_t$ gives
\begin{align}
    \frac{\partial \Phi_t}{\partial \varepsilon}(\varepsilon;s)
    &=
    p_t q(s)-c_t'(\varepsilon)
    +\gamma
    V_{t+1}'\!\left(F(s-q(s)\varepsilon)\right)
    \frac{\partial}{\partial \varepsilon}F(s-q(s)\varepsilon) \\
    &=
    p_t q(s)-c_t'(\varepsilon)
    +\gamma
    V_{t+1}'\!\left(F(s-q(s)\varepsilon)\right)
    F'(s-q(s)\varepsilon)(-q(s))
\end{align}

Setting this derivative equal to zero at $\varepsilon=\varepsilon_t^*(s)$ yields
\[
p_t q(s)-c_t'\!\bigl(\varepsilon_t^*(s)\bigr)
=
\gamma q(s)\,
V_{t+1}'\!\left(F(s-q(s)\varepsilon_t^*(s))\right)
F'(s-q(s)\varepsilon_t^*(s)),
\]
which is \eqref{eq:bellman_foc}.

Now define the post-harvest stock $w_t^*(s) \triangleq s-q(s)\varepsilon_t^*(s)$ and division by $q(s)$ (since $q(s)>0$) gives:
\[
p_t-\frac{c_t'\!\bigl(\varepsilon_t^*(s)\bigr)}{q(s)}
=
\gamma
V_{t+1}'\!\left(F(w_t^*(s))\right)
F'(w_t^*(s)),
\]
which is \eqref{eq:bellman_foc_w}.

Finally, we explain why one should not expect a general closed form. The unknown control enters the first-order condition through the term
\[
V_{t+1}'\!\left(F(s-q(s)\varepsilon)\right),
\]
and, except in special cases, the value function $V_{t+1}$ is itself only defined recursively through the Bellman equation. Therefore the optimizer is characterized implicitly by \eqref{eq:bellman_foc} or \eqref{eq:bellman_foc_w}, but generally cannot be written explicitly in closed form.
\end{proof}



Theorem~\ref{thm:bellman_characterization} establishes that the optimal policy is characterized recursively by the Bellman equation and, in general, need not admit a closed-form expression. We next complement this result with a Hamiltonian analysis. This yields a structural characterization of the optimal control through the switching function and, under additional assumptions, a closed-form expression for the singular candidate.\footnote{Note that we implemented numerical dynamic-programming solutions of the finite-horizon problem which indicate that (for the evaluated states and time periods) the optimal harvesting effort is strictly interior. This is consistent with Theorem~\ref{thm:bellman_characterization} and motivates the explicit treatment of the singular case in Proposition~\ref{prop:hamiltonian_characterization_and_singular}.}

\bigskip
\begin{proposition}[Hamiltonian characterization and singular candidate]
\label{prop:hamiltonian_characterization_and_singular}
Consider the single-resource, single-controller problem over a finite horizon $T$, with state $s_t \in [0,\infty)$, control $\varepsilon_t \in [0,E_{\max}]$, and stock dynamics and one-period cost as described in Section \ref{Fishery Model}, and a linear harvesting cost. Suppose the system remains in the uncapped-harvest region, so that $H(\varepsilon_t,s_t)=q(s_t)\varepsilon_t$. Let $\sigma_t$ denote the switching function of the Hamiltonian. Then the optimal control at time $t+1$ satisfies:
\begin{equation} \label{Eq: controler}
    \varepsilon_{t+1}^* \in
    \begin{cases}
    \{E_{\max}\}, & \text{if } \sigma_t>0,\\[1mm]
    [0,E_{\max}], & \text{if } \sigma_t=0,\\[1mm]
    \{0\}, & \text{if } \sigma_t<0.
    \end{cases}
\end{equation}

\noindent
where $\sigma_t = (\gamma^{t+1}p_{t+1}-\lambda_{t+1})q(F(w_t))-\gamma^{t+1}c$ and $w_t$ is the transformed state, $w_t \triangleq s_t - H(\varepsilon_t, s_t) = s_t-q(s_t)\varepsilon_t$.

When $\sigma_t=0$, the first-order Hamiltonian maximization condition does not determine the control uniquely.

In the special case  where we have fixed operational cost independent of the effort ($c=0$, i.e., $c_t(\varepsilon_t)=c'$) and the optimal trajectory contains a singular arc (in particular, the switching function satisfies $\sigma_t=0$ and $\sigma_{t-1}=0$ on the relevant consecutive time-steps), then the singular branch ($\sigma_t=0$) admits the candidate control\footnote{This is a singular candidate control. Its optimality is not automatic: one must still verify feasibility and compare it against the boundary controls $0$ and $E_{\max}$.}
\[
\varepsilon_t^{\mathrm{sing}}
=
2S^{eq}\left(1-\frac{w_t^{\mathrm{sing}}}{s_t}\right),
\]
where
\begin{equation}
\label{eq:wstar_lambert_combined}
w_t^{\mathrm{sing}}
=
\frac{S^{eq}}{g}
\left[
1-
W\!\left(
\frac{p_t}{\gamma p_{t+1}}e^{\,1-g}
\right)
\right],
\end{equation}
and $W$ denotes the Lambert $W$ function on an admissible branch. 
\end{proposition}

\begin{proof}
To improve readability, we have omitted the subscripts $n$, and $r$, since we consider a setting where a single resource is harvested and sold and which the means of harvesting are controlled by a single entity.

We begin by decoupling the state (i.e., for the purposes of this proof, the current resource stock, $s_t$) and the control ($\varepsilon_t$). Let $w_t$ be the remaining stock after harvest at time-step $t$:
\begin{equation*} \label{Eq: change of state variable - 1}
  w_t \triangleq s_t - H(\varepsilon_t, s_t)
\end{equation*}

Therefore,
\begin{equation} \label{Eq: change of state variable - 2}
  s_{t+1}=w_{t+1}+H(\varepsilon_{t+1},s_{t+1})
\end{equation}

and
\begin{equation} \label{Eq: change of state variable - 2.5}
  s_{t+1}=F(s_t - H(\varepsilon_t,s_t))=F(w_t)
\end{equation}

Combining (\ref{Eq: change of state variable - 2}) and (\ref{Eq: change of state variable - 2.5}), we get:
\begin{equation} \label{Eq: change of state variable - 3}
  w_{t+1}=F(w_t)-H(\varepsilon_{t+1},F(w_t))
\end{equation}

In the uncapped-harvest region, the total harvest can be written as:
\begin{equation*} \label{Eq: Total harvest modified}
  H(\varepsilon_{t+1},F(w_t))=q(F(w_t))\varepsilon_{t+1}
\end{equation*}

and the state dynamics can be written as:
\begin{equation} \label{Eq: change of state variable - 4}
  w_{t+1}=F(w_t)-q(F(w_t))\varepsilon_{t+1}
\end{equation}

We want to maximize the total discounted revenue for a given time horizon $T$. Given the above derivations, the optimization problem can be written as:
\begin{align*} \label{Eq: Optimization Problem 2 - Objective}
  &\underset{\varepsilon_t}{\max} \underset{t = 0}{\overset{T}{\sum}} \gamma^t u_{t}(\varepsilon_{t}, s_t) \\
  &\underset{\varepsilon_t}{\max} \underset{t = 0}{\overset{T}{\sum}} \gamma^t \left( p_t H(\varepsilon_{t}, s_t) - c_t(\varepsilon_{t}) \right) \\
  &\underset{\varepsilon_t}{\max} \underset{t = 0}{\overset{T}{\sum}} \gamma^t \left( p_t q(s_t) \varepsilon_t - (c \varepsilon_{t} + c' )  \right) \\
  &\underset{\varepsilon_{t+1}}{\max} \underset{t = -1}{\overset{T-1}{\sum}} \gamma^{t+1} \left( p_{t+1} q(s_{t+1}) \varepsilon_{t+1} - (c \varepsilon_{t+1} + c' )  \right) \\
  &\underset{\varepsilon_{t+1}}{\max} \underset{t = -1}{\overset{T-1}{\sum}} \gamma^{t+1} \left( p_{t+1} q\left( F(w_t) \right) - c \right) \varepsilon_{t+1} - \gamma^{t+1} c' \\
  &\text{subject to } w_{t+1}=F(w_t)-q(F(w_t))\varepsilon_{t+1}
\end{align*}

\noindent
where we assumed linear cost, $c_t(\varepsilon_t) = c \varepsilon_t + c'$, as in the related literature (e.g.,~\cite{gordon1954economic,de2008sustainable}). Let $w_{-1}=s_0=S^{eq}$. Solving the optimization problem is equivalent to finding the control that optimizes the Hamiltonians \cite{lenhart2007optimal,ding2010introduction_optimalcontrol}. Let $\lambda = (\lambda_{-1}, \lambda_{0}, \dots, \lambda_{T-1})$ denote the adjoint function. The Hamiltonian at time-step $t$ is given by:
\begin{equation} \label{Eq: Hamiltonian}
\begin{split}
  \mathcal H_t &= \gamma^{t+1} \left( p_{t+1} q\left( F(w_t) \right) - c \right) \varepsilon_{t+1} - \gamma^{t+1} c' + \lambda_{t+1}(F(w_t)-q(F(w_t))\varepsilon_{t+1}) \\
  & = \left( \left( \gamma^{t+1} p_{t+1}-\lambda_{t+1} \right)q(F(w_t)) - \gamma^{t+1} c \right)\varepsilon_{t+1} - \gamma^{t+1} c'+ \lambda_{t+1}F(w_t)
\end{split}
\end{equation}

The adjoint equations are given by (see~\cite{ding2010introduction_optimalcontrol}):
\begin{equation*} \label{Eq: Hamiltonian_conditions}
\begin{split}
  \lambda_t & = \frac{\partial \mathcal H_t}{\partial w_t} \\
  \lambda_T & = 0 \\
\end{split}
\end{equation*}

\noindent
where T is the time horizon, and the optimal control maximizes the Hamiltonian over $\varepsilon_{t+1}\in[0,E_{\max}]$ at each time-step.

Therefore the switching function is
\begin{equation}
\label{eq:switching_function_combined_prop_proof}
\sigma_t
=
(\gamma^{t+1}p_{t+1}-\lambda_{t+1})q(F(w_t))-\gamma^{t+1}c.
\end{equation}

For fixed $w_t$ and $\lambda_{t+1}$, the Hamiltonian is affine in the control:
\[
\mathcal H_t
=
\sigma_t \varepsilon_{t+1}
+\Bigl(-\gamma^{t+1}c'+\lambda_{t+1}F(w_t)\Bigr).
\]
Since the admissible control set is the compact interval $[0,E_{\max}]$, maximizing $\mathcal H_t$ over $\varepsilon_{t+1}\in[0,E_{\max}]$ is immediate:
\begin{itemize}
    \item if $\sigma_t>0$, then $\mathcal H_t$ is strictly increasing in $\varepsilon_{t+1}$, so the unique maximizer is $\varepsilon_{t+1}^*=E_{\max}$;
    \item if $\sigma_t<0$, then $\mathcal H_t$ is strictly decreasing in $\varepsilon_{t+1}$, so the unique maximizer is $\varepsilon_{t+1}^*=0$;
    \item if $\sigma_t=0$, then $\mathcal H_t$ is constant in $\varepsilon_{t+1}$, so every control in $[0,E_{\max}]$ satisfies the first-order Hamiltonian maximization condition.
\end{itemize}
This proves the first part of the proposition.

We now analyze the singular case under the additional assumptions listed in the statement. Specifically, we assume fixed operational cost independent of the effort ($c=0$, i.e., $c_t(\varepsilon_t)=c'$) and that the optimal trajectory contains a singular arc (in particular, the switching function satisfies $\sigma_t=0$ and $\sigma_{t-1}=0$ on the relevant consecutive time-steps).

Since $c=0$, the switching function becomes
\[
\sigma_t
=
(\gamma^{t+1}p_{t+1}-\lambda_{t+1})q(F(w_t)).
\]
On a singular arc, $\sigma_t=0$. Because $q(F(w_t))>0$ in the relevant region, it follows that
\begin{equation}
\label{eq:lambda_singular_tplus1_rewrite}
\lambda_{t+1}=\gamma^{t+1}p_{t+1}.
\end{equation}

Differentiating the Hamiltonian with respect to $w_t$ and using the first adjoint equation ($\lambda_t=\frac{\partial \mathcal H_t}{\partial w_t}$) we get:
\[
\lambda_t
=
(\gamma^{t+1}p_{t+1}-\lambda_{t+1})q'(F(w_t))F'(w_t)\varepsilon_{t+1}
+\lambda_{t+1}F'(w_t).
\]
Using \eqref{eq:lambda_singular_tplus1_rewrite}, the first term vanishes, and therefore
\begin{equation}
\label{eq:adjoint_singular_rewrite}
\lambda_t=\lambda_{t+1}F'(w_t).
\end{equation}

Since the singular arc persists at consecutive times, we can perform a change of variables $t = t' - 1$ in Equation \ref{eq:lambda_singular_tplus1_rewrite} and get
\[
\lambda_t=\gamma^t p_t.
\]
Combining this with \eqref{eq:lambda_singular_tplus1_rewrite} and \eqref{eq:adjoint_singular_rewrite} yields
\[
\gamma^t p_t
=
\lambda_t
=
\lambda_{t+1}F'(w_t)
=
\gamma^{t+1}p_{t+1}F'(w_t),
\]
hence
\begin{equation}
\label{eq:Fprime_singular_rewrite}
F'(w_t)=\frac{p_t}{\gamma p_{t+1}}.
\end{equation}

We now solve this equation for the growth function of our model (Equation \ref{Eq: Spawner-recruit function}). Differentiating Equation \ref{Eq: Spawner-recruit function} (for brevity, write $S \triangleq S^{eq}$), we have:
\[
F'(x)
=
\exp\!\left(g\left(1-\frac{x}{S}\right)\right)
\left(1-\frac{gx}{S}\right).
\]
Substituting this into \eqref{eq:Fprime_singular_rewrite} gives
\[
\exp\!\left(g\left(1-\frac{w_t}{S}\right)\right)
\left(1-\frac{gw_t}{S}\right)
=
\frac{p_t}{\gamma p_{t+1}}.
\]
Let
\[
y\triangleq\frac{p_t}{\gamma p_{t+1}}
\qquad\text{and}\qquad
z\triangleq1-\frac{gw_t}{S}.
\]
Then
\[
w_t=\frac{S}{g}(1-z).
\]
Also,
\[
g\left(1-\frac{w_t}{S}\right)
=
g\left(1-\frac{1-z}{g}\right)
=
g-1+z.
\]
Therefore the singularity condition becomes
\[
e^{g-1+z}z=y,
\]
or equivalently (using the Lambert $W$ function),
\[
z=W\!\left(ye^{\,1-g}\right).
\]
Substituting back $y=\frac{p_t}{\gamma p_{t+1}}$ and $z\triangleq1-\frac{gw_t}{S}$ yields
\[
w_t^{\mathrm{sing}}
=
\frac{S^{eq}}{g}
\left[
1-
W\!\left(
\frac{p_t}{\gamma p_{t+1}}e^{\,1-g}
\right)
\right].
\]

Finally, by definition of the transformed state, $w_t \triangleq s_t - H(\varepsilon_t, s_t) = s_t-q(s_t)\varepsilon_t$, the corresponding singular candidate control is
\[
\varepsilon_t^{\mathrm{sing}}
=
\frac{s_t-w_t^{\mathrm{sing}}}{q(s_t)}.
\]
In the linear-catchability region,
\[
q(s_t)=\frac{s_t}{2S^{eq}},
\]
and therefore
\[
\varepsilon_t^{\mathrm{sing}}
=
\frac{s_t-w_t^{\mathrm{sing}}}{s_t/(2S^{eq})}
=
2S^{eq}\left(1-\frac{w_t^{\mathrm{sing}}}{s_t}\right).
\]

This yields the stated singular candidate.\footnote{Note that its optimality is not automatic: to conclude that it is optimal, one must still verify that it is feasible (lies in $[0,E_{\max}]$), that it is consistent with the assumptions under which it was derived (in particular the uncapped-harvest and linear-catchability conditions), and that it yields no lower value than the boundary controls $0$ and $E_{\max}$.}
\end{proof}


From Equation \ref{Eq: controler} of Proposition~\ref{prop:hamiltonian_characterization_and_singular}, under fixed operational cost ($c=0$), maximal-effort harvesting is optimal whenever the switching function is strictly positive. Intuitively, this occurs when the current price $p_t$ is sufficiently large. In other words, as long as the price $p_{t}$ at which the good is sold in each round is large enough, the harvesters exert maximum effort in each round, which can lead to depletion of the resource (we quantify the time needed for that to happen later in the section).


Now let us assume that, in the absence of any policymaker, the market equilibrium (i.e., the free-market outcome) is computed at each time-step. For simplicity, we continue to assume fixed operational cost $c_t(\varepsilon_t)=c'$, as in \cite{10.5555/3463952.3464003}. By the clearing condition of the Fisher market \cite{jain2010eisenberg,mas1995microeconomic}, it holds that at the equilibrium price $p_{t}=p$ each buyer spends its entire budget, i.e., $x_b\cdot p = \beta_b$, where $x_b$ is the amount of the resource allocated to buyer $b$. By summing over all buyers, and assuming for simplicity that the total supply of the resource is $1$, we obtain that $\beta_t =\sum_{b}\beta_{b,t} = p_t$, that is, the equilibrium price is equal to the cumulative budget of all buyers.


Therefore, if the cumulative budget is sufficiently large relative to future budgets, the switching function in Equation \ref{Eq: controler} of Proposition~\ref{prop:hamiltonian_characterization_and_singular} remains strictly positive and maximal-effort harvesting is optimal at every time-step. In other words, the harvesters will harvest with maximum effort at every time-step, potentially depleting the resource. In what follows, we quantify what ``sufficiently large'' means by providing a sufficient condition on the relation between cumulative budgets across time. This is captured in the following theorem.

\bigskip
\begin{theorem}[Sufficient budget condition for maximal-effort harvesting]
\label{thm:budget_bound_sufficient}
For $|\mathcal{R}| = |\mathcal{N}| = 1$, fixed operational cost $c_t(\varepsilon_t)=c'$, and the model dynamics described in Section~\ref{Models}, suppose that the system remains in the uncapped-harvest region and that the Fisher market is cleared at each time-step, so that $p_t=\beta_t\triangleq\sum_b \beta_{b,t}$ when the total supply of the resource is normalized to $1$. If $\beta_T > 0$ and, for every $t<T$,
\begin{equation}
\label{eq:budget_ratio_sufficient_new}
\frac{\beta_t}{\underset{i \in [t+1 \dots T]}{\max} \beta_i}
>
\frac{1}{2S^{eq}}
\sum_{j=1}^{T-t} (\gamma e^g)^j,
\end{equation}
then the maximal effort control $\varepsilon_t^*=E_{\max}$ is optimal for every $t=0,\dots,T$.
\end{theorem}

\begin{proof}
By Proposition~\ref{prop:hamiltonian_characterization_and_singular}, in the present fixed operational cost case $c_t(\varepsilon_t)=c'$, a sufficient condition for maximal-effort harvesting at time $t$ is that the switching function (Equation \ref{eq:switching_function_combined_prop_proof}) be strictly positive. Since $c=0$, the switching function reduces to
\[
(\gamma^t p_t-\lambda_t)q(F(w_{t-1})),
\]
and because by definition $q(F(w_{t-1}))\ge 0$, it is enough to require
\begin{equation}
\label{eq:sufficient_sigma_positive}
\gamma^t p_t > \lambda_t .
\end{equation}
We therefore derive an upper bound on $\lambda_t$.

To improve readability, we omit the subscripts $n$ and $r$, since we consider a setting where a single resource is harvested and sold and which the means of harvesting are controlled by a single entity.

First, we calculate two auxiliary partial derivatives that will be used later on (in the proof by induction). In what follows, we assume that the remaining stock after harvest, $w_t$, is a fraction of the stock equilibrium population, i.e., $w_t = \alpha S^{eq}, \alpha \in [0, 1]$. We have that:
\begin{equation} \label{Eq: theta F}
  \frac{\partial F(w_t)}{\partial w_t} = e^{g \left( 1-\frac{w_t}{S^{eq}} \right)} \left( 1-\frac{g w_t}{S^{eq}} \right)  = e^{g (1 - \alpha)} (1 - \alpha g)  \leq e^{g}, \quad \text{since growth rate $g \in [0, \infty)$}
\end{equation}

\noindent
and, since $q'(x)\le \frac{1}{2S^{eq}}$ for all relevant $x$,
\begin{equation} \label{Eq: theta qF}
  \frac{\partial q\left(F(w_t)\right)}{\partial w_t} 
  = 
  q'\left(F(w_t)\right)\frac{\partial F(w_t)}{\partial w_t}
  \leq
  \frac{1}{2 S^{eq}} \frac{\partial F(w_t)}{\partial w_t}
  =
  \frac{e^{g \left( 1-\frac{w_t}{S^{eq}} \right)} \left( 1-\frac{g w_t}{S^{eq}} \right)}{2 S^{eq}}  \leq \frac{e^{g}}{2 S^{eq}}
\end{equation}

\noindent
The Hamiltonian at time-step $t$ is given by (see Equation \ref{Eq: Hamiltonian}):
\begin{equation*} \label{Eq: Hamiltonian 1}
  \mathcal H_t = \gamma^{t+1} p_{t+1} \varepsilon_{t+1} q(F(w_t)) - \lambda_{t+1} \varepsilon_{t+1} q(F(w_t)) + \lambda_{t+1} F(w_t) + \kappa
\end{equation*}

\noindent
where $\kappa = -\gamma^{t+1} \varepsilon_{t+1} c - \gamma^{t+1} c'$ (in the present theorem $c=0$, so $\kappa=-\gamma^{t+1}c'$).

Next, we show by induction that:
\begin{equation} \label{Eq: to prove - induction}
    \underset{j = 1}{\overset{t}{\sum}} \gamma^{T-(t-j)} p_{T-(t-j)} \frac{e^{j g}}{2 S^{eq}} \geq \lambda_{T - t}
\end{equation}

\noindent
for $1 \leq t \leq T$ where $T$ is the time horizon. For simplicity, we normalize effort so that $E_{\max}=1$.

\paragraph{Base Case:}

For $t = T, \lambda_T = 0$ and $\varepsilon_T\le E_{\max}$, thus for $T = T-1$:
\begin{equation*}
    \lambda_{T-1}  = \frac{\partial \mathcal{H}_{T-1}}{\partial w_{T-1}} = \gamma^{T} p_{T} \varepsilon_{T} \frac{\partial q\left(F(w_{T-1})\right)}{\partial w_{T-1}} \leq \gamma^{T} p_{T} \frac{e^{g}}{2 S^{eq}}
\end{equation*}
\noindent
The base case holds.

\paragraph{Induction Step:}

Let $\lambda_{T-i} \leq \underset{j = 1}{\overset{i}{\sum}} \gamma^{T-(i-j)} p_{T-(i-j)} \frac{e^{j g}}{2 S^{eq}}, \forall i < t$. For $t$ we have that:
\begin{equation*}
\begin{split}
    \lambda_{T-t}  &= \frac{\partial \mathcal{H}_{T-t}}{\partial w_{T-t}} \\
    &= \gamma^{T-t+1} p_{T-t+1} \varepsilon_{T-t+1} \frac{\partial q\left(F(w_{T-t})\right)}{\partial w_{T-t}} \\
    &- \lambda_{T-t+1} \varepsilon_{T-t+1} \frac{\partial q\left(F(w_{T-t})\right)}{\partial w_{T-t}} + \lambda_{T-t+1} \frac{\partial F(w_{T-t})}{\partial w_{T-t}} \\
    & \leq \gamma^{T-(t-1)} p_{T-(t-1)} \frac{\partial q\left(F(w_{T-t})\right)}{\partial w_{T-t}} + \lambda_{T-(t-1)} \frac{\partial F(w_{T-t})}{\partial w_{T-t}} \\
    & \leq \gamma^{T-(t-1)} p_{T-(t-1)} \frac{e^{g}}{2 S^{eq}} + e^{g} \underset{j = 1}{\overset{t-1}{\sum}} \gamma^{T-(t-1-j)} p_{T-(t-1-j)} \frac{e^{j g}}{2 S^{eq}} \\
    & = \underset{j = 1}{\overset{t}{\sum}} \gamma^{T-(t-j)} p_{T-(t-j)} \frac{e^{j g}}{2 S^{eq}}
\end{split}
\end{equation*}

That is, (\ref{Eq: to prove - induction}) holds.

To improve readability, we will perform a change of variables in (\ref{Eq: to prove - induction}), leading to:
\begin{equation} \label{Eq: bound lambda}
    \underset{j = 1}{\overset{T-t}{\sum}} \gamma^{t+j} p_{t+j} \frac{e^{j g}}{2 S^{eq}} \geq \lambda_{t}
\end{equation}

\noindent
for $t = 0, \dots, T-1$.

Plugging in Equation \ref{eq:sufficient_sigma_positive}, we have that to harvest at maximum effort, a sufficient condition is:
\begin{equation} \label{Eq: bound on prices}
\begin{split}
    \gamma^t p_t &> \underset{j = 1}{\overset{T-t}{\sum}} \gamma^{t+j} p_{t+j} \frac{e^{j g}}{2 S^{eq}} \geq \lambda_t \\
    p_t &> \frac{1}{2 S^{eq}} \underset{j = 1}{\overset{T-t}{\sum}} \gamma^{j} p_{t+j} e^{j g} \\
    p_t > \frac{\max_{i \in [t+1 \dots T]} p_i}{2 S^{eq}} \underset{j = 1}{\overset{T-t}{\sum}} \left( \gamma e^{g} \right)^j &> \frac{1}{2 S^{eq}} \underset{j = 1}{\overset{T-t}{\sum}} \gamma^{j} p_{t+j} e^{j g}
\end{split}
\end{equation}

Finally, by market clearing in the normalized single-good Fisher market, the equilibrium price equals the cumulative budget:
\[
p_t=\beta_t \triangleq \sum_b \beta_{b,t}.
\]

Therefore a sufficient condition for maximal-effort harvesting at every time-step is
\[
\frac{\beta_t}{\max_{i\in[t+1,\dots,T]} \beta_i}
>
\frac{1}{2S^{eq}}
\sum_{j=1}^{T-t}(\gamma e^g)^j,
\qquad \forall t<T,
\]
which is exactly \eqref{eq:budget_ratio_sufficient_new}. The conclusion then follows from Proposition~\ref{prop:hamiltonian_characterization_and_singular}.
\end{proof}


We remark that this is only a \emph{sufficient} condition, as we have applied several simplifications to the derivation to end up with a formula from which useful information can be extracted. Computing a necessary condition appears to require extremely tedious derivations, as the bound involves computing partial derivatives of a recursive expression which results in a function composed over $T$ time-steps. Still, as we demonstrate in the following, this sufficient condition is enough to support our claim that depletion can realistically happen in environments of interest. 

According to Theorem~\ref{thm:budget_bound_sufficient}, and in particular condition~\ref{eq:budget_ratio_sufficient_new}, as we reach the time horizon $T$, a smaller budget is sufficient to prompt the harvester into harvesting with maximum effort\footnote{More precisely, keep the switching function in the strictly positive region and hence make maximal-effort harvesting optimal, provided the system remains in the regime where the theorem applies.}. As a grounding example, consider the following concrete parameters of the model: $t = 0$ (which corresponds to the worst case), $\gamma = 0.9$, $g = -\ln(\gamma) \approx 0.105$, a time horizon of 5 years ($T = 60$, assuming a time-step is one month), and $S^{eq} = M_s K N = 0.8 \times 0.79 \times 1$ (see Section \ref{Resources}). In this case, (\ref{eq:budget_ratio_sufficient_new}) results in:
\begin{equation} \label{example bound}
  \frac{\beta_0}{\underset{i \in [1 \dots T]}{\max} \beta_i}  > \frac{1}{2 S^{eq}} \underset{j = 1}{\overset{T}{\sum}} \left( \gamma e^{g} \right)^j  = \frac{T}{2 S^{eq}} \approx 47.5
\end{equation}

This indicates that there exist environments in which the \emph{cumulative} budget required to sustain maximal-effort incentives in every time-step is of the same order of magnitude as its initial value. Given that $\beta$ is itself a cumulative measure, such environments are reasonable since, even if the budget for each buyer is (almost) fixed, the number of buyers that actually appear in the market might decrease over time.

Let us now study the stock dynamics under the maximal-effort policy itself.
Let $D(s) = F(s - H(E_{\max},s))$, where $F(\cdot)$ and $H(\cdot)$ are given by (\ref{Eq: Spawner-recruit function}) and (\ref{Eq: Total harvest}), respectively. The resource will get depleted in $t_d \in \mathbb{Z}^+$ time-steps, such that $D^{t_d}(s_0) < \delta S^{eq}$, i.e., when the stock drops below a percentage of the equilibrium population.\footnote{$D^x(\cdot)$ denotes the $x$th function composition of $D(\cdot)$ with itself.} For $\delta = 10^{-3}$ and the same parameters specified in the example above, we obtain $t_d = 5$ (months). For $\delta = 10^{-6}$, we obtain $t_d = 10$ (months). Both are considerably lower than the desired time horizon ($T = 60$ months). In simple terms, the practical example of this section comes to show that there exist environments in which:

\begin{quote}\centering
    \emph{Market-equilibrium prices can sustain strong incentives for aggressive harvesting, which can lead to resource depletion even under optimal harvesting.}
\end{quote}


Another takeaway from the results of this section is that computing the optimal harvesting strategy analytically is quite challenging, as there is no closed form expression and the $\lambda_t$ values of Proposition~\ref{prop:hamiltonian_characterization_and_singular} can only be computed recursively via repeated partial derivation. Similarly, preventing the potential depletion of the resource requires the policy-maker to at the very least have knowledge of the maximum and minimum budgets over the time horizon $T$, which typically will not be the case in practice. This motivates the use of DRL over the analytical approach, as we also advocate in the Introduction.

\subsection{Simulation Settings: Fishery \& Market} \label{Common Fishery Settings}

\subsubsection{Resources} \label{Resources}

We simulated two scenarios, one with more plentiful resources, and a \emph{scarce resources} scenario, using the findings of \cite{10.5555/3463952.3464003} as a guide on the selection of the proper $S_r^{eq}$ values. Specifically, we set $S_r^{eq} = M_s K N$, where $K = (e^{g_r} \Phi_{\max}) / (2 (e^{g_r} - 1)) \approx 0.79$ is a constant, and $M_s \in \mathbb{R}^+$ is a multiplier that adjusts the scarcity of the resource (difficulty of the problem). \footnote{For example, for $M_s = 1$ the resource will not get depleted, even if all agents harvest at maximum effort. Yet, coordination is far from trivial; see \cite{10.5555/3463952.3464003}.} For the majority of the simulations we used $M_s = 0.8$, while for the \emph{scarce resources} scenario we used $M_s = 0.45$.\footnote{In the simulations of \cite{10.5555/3463952.3464003}, for $N = 8$ and $M_s \leq 0.4$, the agents failed to find a sustainable strategy, and always depleted the resource.} We set the maximum effort at $\Phi_{\max} = 1$, the growth rate at $g_r = 1$, and the initial population at $s_0 = S_r^{eq}$ (i.e., the stock starts from the equilibrium population), $\forall r \in \mathcal{R}$.

\subsubsection{Harvesters} \label{Harvesters Settings}

We set the skill level $\eta_{n, r} = 0.5$ for all agents and resources, except for one resource for each agent, specifically $\eta_{n, r} = 1$ if $n = r$. Finally, we assume no cost in harvesting, i.e., $c_{n, r, t}(\cdot) = 0$, $\forall n \in \mathcal{N}$, $\forall t$.

\subsubsection{Buyers} \label{Buyers Settings}

Every time-step, a new set of buyers appears at the market, with budgets and valuations drawn uniformly at random on $[0, 1]$. While we will be referring to individual buyers throughout the text, our analysis extends trivially to the case where each buyer represents a \emph{class} of buyers with similar budgets and valuations. The assumption that buyers in a market appear in groups with common characteristics is common in both theory and practice, and it is in fact the basis of the well-established \emph{market segmentation} approach \cite{wedel2000market}.

\section{Agent Architecture} \label{Agent Architecture}

\subsection{Decentralised Multi-Agent Deep RL} \label{Multi-Agent Reinforcement Learning}

We consider a \emph{decentralized} multi-agent reinforcement learning scenario in a partially observable general-sum stochastic game (e.g.,~\cite{10.5555/3091574.3091594,Shapley1095}). At each time-step, agents take actions based on a partial observation of the state space, and receive an individual reward. Each agent learns a policy independently, using a two-layer ($64$ neurons each) feed-forward neural network for the policy approximation. More formally, let $\mathcal{N} = \{1, \dots, N\}$ denote the set of agents, and $\mathcal{M}$ be an $N$-player, partially observable stochastic game defined on a set of states $\mathcal{S}$. An observation function $\mathcal{O}^n: \mathcal{S} \rightarrow \mathds{R}^d$ specifies agent $n$'s $d$-dimensional view of the state space. Let $\mathcal{A}^n$ denote the set of actions for agent $n \in \mathcal{N}$, and $\bm{a} = \times_{\forall n \in \mathcal{N}} a^n$, where $a^n \in \mathcal{A}^n$, the joint action. The states change according to a transition function $\mathcal{T}: \mathcal{S} \times \mathcal{A}^1 \times \dots \times \mathcal{A}^N \rightarrow \Delta(\mathcal{S})$, where $\Delta(\mathcal{S})$ denotes the set of probability distributions over $\mathcal{S}$. Every agent $n$ receives an individual reward based on the current state $\xi_t \in \mathcal{S}$ and joint action $\bm{a}_t$. The latter is given by the reward function $r^n : \mathcal{S} \times \mathcal{A}^1 \times \dots \times \mathcal{A}^N \rightarrow \mathds{R}$. Finally, each agent learns a policy $\pi ^n : \mathcal{O}^n \rightarrow \Delta(\mathcal{A}^n)$ independently through their own experience of the environment (observations and rewards). Let $\bm{\pi} = \times_{\forall n \in \mathcal{N}} \pi^n$ denote the joint policy. The goal for each agent is to maximize the long term discounted payoff, as given by $V^n_{\bm{\pi}} (\xi_0) = \mathds{E} \left[ \sum_{t = 0}^\infty \gamma^t r^n(\xi_t, \bm{a}_t) | \bm{a}_t \sim \bm{\pi}_t, \xi_{t+1} \sim \mathcal{T}(\xi_t, \bm{a}_t) \right]$, where $\gamma$ is the discount factor and $\xi_0$ is the initial state.

\subsubsection{Learning Algorithm} \label{Learning Algorithm}

The policies for all learning agents (harvesters and the policymaker)\footnote{The buyers are not learning agents; see Section \ref{Buyer Architecture}.} are trained using the Proximal Policy Optimization (PPO) algorithm \cite{DBLP:journals/corr/SchulmanWDRK17}. PPO was chosen because it avoids large policy updates, ensuring a smoother training, and avoiding catastrophic failures. Please see Appendix \ref{supplement: Agent Architecture} for implementation details and hyperparameters.

\subsection{Harvesters' Architecture} \label{Harvester Agent Architecture}

Each harvester's input (observation) is a tuple $\langle \bm{p}_{t-1}, \bm{\phi}_{n, t-1}, \bm{u}_{n, t-1}(\cdot) \rangle$ consisting of the vector of prices for the resources, the vector of individual effort exerted for every resource, and reward (cumulative out of all the resources) obtained in the previous time-step. The output is a vector of continuous action values $a_t^n = \bm{\phi}_{n, t} \in [0, \Phi_{\max}]^{|\mathcal{R}|}$ specifying the current effort level to exert for harvesting each resource. The reward received from the environment corresponds to the revenue, i.e., $\sum_{r \in \mathcal{R}} u_{n, r, t}(\varepsilon_{n, r, t},s_{r, t})$.

\subsection{Policymaker Architecture} \label{Policymaker Agent Architecture}

The input of the policymaker is a tuple $\langle \bm{\varepsilon}_t, \bm{s}_t, \bm{\beta_t}, G(\bm{v}_t) \rangle$, where $\bm{\varepsilon}_t$ is the efforts exerted by all the harvesters for all the resources, $\bm{s}_t$ is the current stock level of each resource, $\bm{\beta_t}$ is the budgets of the current set of buyers (recall that a random set of buyers appears at the market at each time-step), and finally, $G(\bm{v}_t)$ are the obfuscated valuations of the buyers, obfuscated by a function $G(\cdot)$. The output is a vector of continuous action values $a_t = \bm{p}_{t} \in [0, \infty]^{|\mathcal{R}|}$ that corresponds to the prices.

\subsubsection{Multi-objective Optimization} \label{Multi-objective Optimization}

The policymaker's reward is the weighted average of the desired objectives, specifically:
\begin{align} \label{Eq: policymaker's reward}
\begin{split}
  w_{h} \frac{1}{|\mathcal{N}|} \sum_{n \in \mathcal{N}} \bm{u}_{n, t}(\cdot) + w_{b} \frac{1}{|\mathcal{B}|} \sum_{b \in \mathcal{B}} \bm{u}_{b, t}(\cdot) + \\
  w_{s} \min_{r \in \mathcal{R}} \left( \min( s_{r, t} - S_{r}^{eq}, 0) \right) + w_{f} Fair(\mathbf{x})
\end{split}
\end{align}

\noindent
where $w_{h}$, $w_{b}$, $w_{s}$, $w_{f}$ $\in [0, 1]$ correspond to the weights for the harvesters' social welfare objective (sum of utilities, $\sum_{n \in \mathcal{N}} \bm{u}_{n, t}(\cdot)$), the buyers' social welfare objective (sum of utilities, $\sum_{b \in \mathcal{B}} \bm{u}_{b, t}(\cdot)$), the sustainability objective (defined in this work as the maximum negative deviation from the equilibrium stock, $\underset{r \in \mathcal{R}}{\min}\left( \min( s_{r, t} - S_{r}^{eq}, 0) \right)$), and the fairness objective ($Fair(\mathbf{x})$). Given the broad literature on fairness, we evaluated three different well-established fairness indices: the Jain index \cite{DBLP:journals/corr/cs-NI-9809099}, the Gini coefficient \cite{gini1912variabilita}, and the Atkinson index \cite{ATKINSON1970244}.\footnote{For brevity, we only report results on the Jain index. Similar results were obtained for the other indices (see Appendix \ref{supplement: Additional Simulation Results} ).} It is important to note that the proposed technique is not limited to our choice of objectives; rather it can be used for \emph{any combination of objectives}.

\subsection{Robustness} \label{Robustness}

\subsubsection{Valuation Obfuscation} \label{Valuation Obfuscation}

To test the robustness of our policymaker in more realistic scenarios, we considered the case where the buyer's valuations are \emph{obfuscated}. To put this into context, note that one of the idealized assumptions that allows the market equilibrium to be computed centrally is that all the information of the market is \emph{completely and accurately} known. For good supplies and budgets, this assumption is reasonable, as these are typically observable or inferable, and qualify as ``hard'' information \cite{liberti2019information} (see also~\cite{branzei2019walrasian}). In contrast, the valuations of the agents are ``soft'' information; they are hard to elicit, since they are expressed on a cardinal scale, and are possibly even accurately unknown to the agents themselves. The literature on computational social choice theory \cite{brandt2016handbook} has been concerned with the effect of limited or noisy valuation information on the desired outcomes of a system.  

We considered three different obfuscation functions for the buyers' valuations: (i) the identity function $G(x) = x$ (no obfuscation) -- which we used in the majority of the simulations -- (ii) a function that splits $[0, 1]$ into $k$ bins, and each valuation value is replaced by the midpoint of the bin interval (average value of the endpoints), and (iii) a function that adds uniform noise on $(-y, y)$, i.e., $G(x, y) = \operatorname{clip} \left( x + \mathcal{U}(-y, y), 0, 1  \right)$. 

The bins approach corresponds to the case where the agents are not asked to provide accurate cardinal values, but instead they provide scores that somehow encode their actual values. As the literature of the distortion in computational social choice suggests, such an elicitation device is cognitively much more conceivable (see \cite{anshelevich2021distortion} and references therein). The added noise approach corresponds to the case where agents are uncertain about their own values, so they end up reporting noisy estimates of their true value. This approach is clearly reminiscent of the literature on noisy estimates of ground truth, pioneered by \cite{mallows1957non} but in fact dating back to the works of Marquis de Condorcet, more than two centuries ago.

\subsubsection{Effort Misestimation} \label{Valuation Misestimation}

We also test the robustness of the policymaker against harvesters that misestimate their exerted effort, by adding uniform noise $\mathcal{U}(-y, y)$ to the effort values. This can be considered as a form of robustness against sim-to-real gap. In simple terms, agents would attempt to fish with some effort $\varepsilon$, but due to hardware implementation mismatching, partial observability, noisy environments etc., they might experience perturbations in the actual effort exerted, now $\varepsilon' = \operatorname{clip} \left( \varepsilon + \mathcal{U}(-y, y), 0, 1 \right)$.

\subsection{Buyers' Architecture} \label{Buyer Architecture}

The buyers are not learning agents; rather, they maximize their utility. Every time-step, a new set of buyers appears at the market, with budgets and valuations drawn uniformly at random on $[0, 1]$. To allocate resources to buyers (for the cases where the price vector $\bm{p}$ is computed by the policymaker, i.e., the non-market equilibrium scenario), we solve the constraint optimization problem of (\ref{Eq: constraint optimization to allocate resources to buyers}) that assigns each buyer a bundle based on their budget constraints, aiming to maximize the social welfare of the buyers:
\begin{eqnarray} \label{Eq: constraint optimization to allocate resources to buyers}
  \max \ \ & & \sum_{b \in \mathcal{B}} u_b(\mathbf{x}) \\
  s.t. \ \ & & u_b(\mathbf{x}) = \sum_{r \in \mathcal{R}} x_{b, r} v_{b, r},\ \ \forall \ b \in \mathcal{B} \nonumber \\
  & & \sum_{r \in \mathcal{R}} x_{b, r} p_{r} \leq \beta_b,\ \ \forall \ b \in \mathcal{B} \nonumber \\
  & & \sum_{b \in \mathcal{B}} x_{b, r} \leq e_r,\ \ \forall \ r \in \mathcal{R} \nonumber \\
  & & x_{b, r}\geq 0,\ \ \forall \ b \in \mathcal{B}, \ \ r \in \mathcal{R}
  \nonumber
\end{eqnarray}

\subsection{Scalability} \label{Scalability}

We simulated an environment with 8 harvesters, 8 buyer classes, and 4 resources ($N = 8, R = 4, B = 8$, where we overload $B$ to denote the number of \emph{classes} of buyers).

Our approach can scale gracefully to \emph{arbitrarily large} number of harvesters and buyers. This is because the policymaker observes the cumulative harvest per resource (not individual harvest, see Section \ref{Policymaker Agent Architecture}) and it is common practice to split buyers into classes (see Section \ref{Buyers Settings}) with similar budgets and valuations. Thus, assuming that the number (types) of resources and buyer classes stays the same, the size/capacity of the policymaker's network does not need to grow. For completeness, we also simulated a larger scale scenario ($N = 12, R = 6, B = 12$) and achieved similar results (see Appendix \ref{supplement: Additional Simulation Results}).

\begin{figure*}[t!]
  \centering
  \begin{subfigure}[t]{0.48\linewidth}
    \centering
    \includegraphics[width = 1 \linewidth, trim={0.7em 0.7em 0.7em 0.7em}, clip]{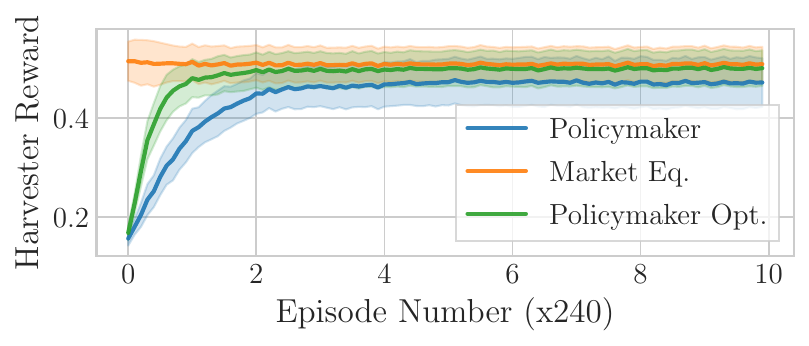}
    \caption{}
    \label{fig: harvester_reward}
  \end{subfigure}
  ~ 
  \begin{subfigure}[t]{0.48\textwidth}
    \centering
    \includegraphics[width = 1 \linewidth, trim={0.7em 0.7em 0.7em 0.7em}, clip]{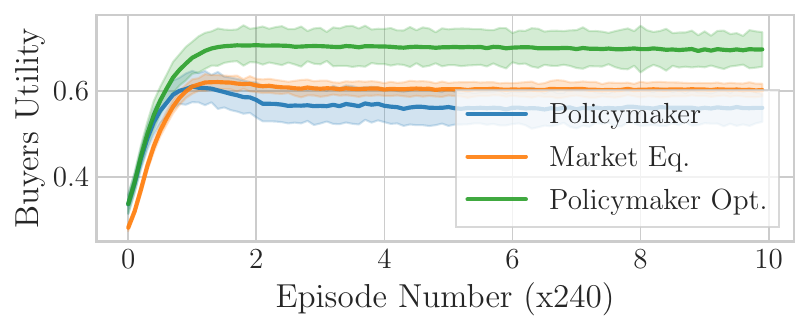}
    \caption{}
    \label{fig: buyers_utility}
  \end{subfigure}
  
  \begin{subfigure}[t]{0.48\textwidth}
    \centering
    \includegraphics[width = 1 \linewidth, trim={0.7em 0.7em 0.7em 0.7em}, clip]{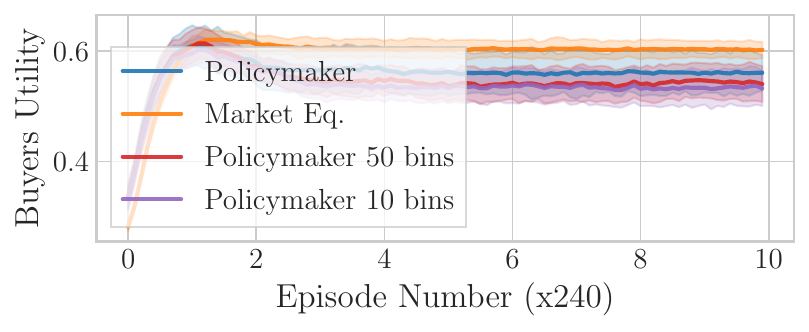}
    \caption{}
    \label{fig: buyers_utility_noisy_valuations_bins}
  \end{subfigure}
  ~
  \begin{subfigure}[t]{0.48\textwidth}
    \centering
    \includegraphics[width = 1 \linewidth, trim={0.7em 0.7em 0.7em 0.7em}, clip]{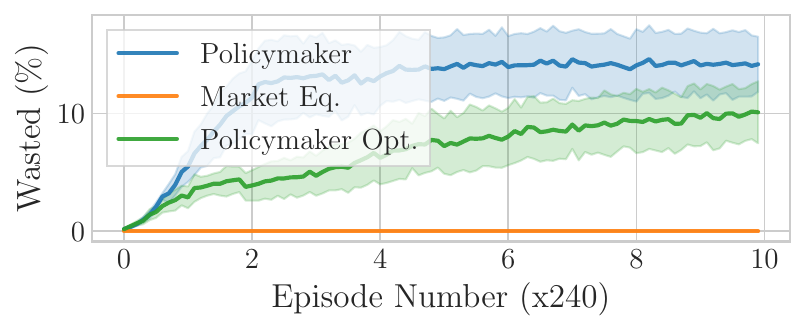}
    \caption{}
    \label{fig: wasted}
  \end{subfigure}
  
  \begin{subfigure}[t]{0.48\textwidth}
    \centering
    \includegraphics[width = 1 \linewidth, trim={0.7em 0.7em 0.7em 0.7em}, clip]{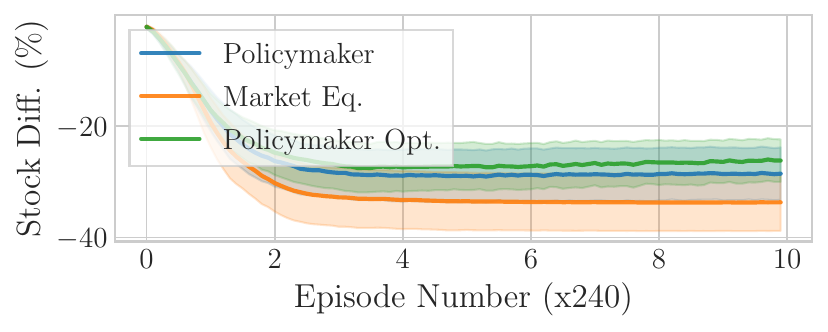}
    \caption{}
    \label{fig: stock_difference}
  \end{subfigure}
  ~ 
  \begin{subfigure}[t]{0.48\textwidth}
    \centering
    \includegraphics[width = 1 \linewidth, trim={0.7em 0.7em 0.7em 0.7em}, clip]{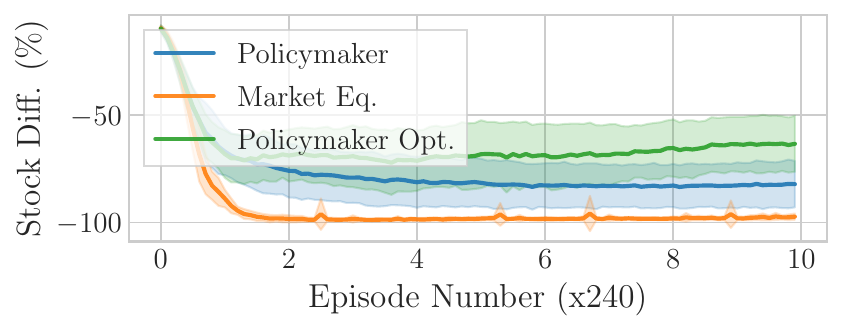}
    \caption{}
    \label{fig: stock_difference_scarce}
  \end{subfigure}%
  \caption{Evolution of several metrics over the number of training episodes. The orange line is the baseline (market equilibrium prices). The blue line refers to the vanilla policymaker where each objective in the reward function has the same weight (see Section \ref{Results}). The green line refers to the policymaker that only optimizes the specific objective of each figure (i.e., in \ref{fig: harvester_reward} we set $w_{h} = 1$, in \ref{fig: buyers_utility} we set $w_{b} = 1$, and in \ref{fig: wasted}, \ref{fig: stock_difference}, and \ref{fig: stock_difference_scarce} we set $w_{s} = 1$ and the rest of the weights to 0). The red and purple lines in \ref{fig: buyers_utility_noisy_valuations_bins} refer to a policymaker with obfuscated valuations (see Section \ref{Valuation Obfuscation}). In \ref{fig: stock_difference_scarce} we have a scarce resource setting (see Section \ref{Common Fishery Settings}). Shaded areas represent one standard deviation.}
  \label{fig: Results}
\end{figure*}

\section{Simulation Results} \label{Results}

We study the effect -- with regard to diverse objectives such as sustainability and resource wastefulness, fairness, buyers’ and sellers’ welfare, etc. -- of introducing the proposed policymaker to our complex economic system, compared to having the market equilibrium prices (as given by solving the convex optimization program we describe in Section \ref{sec:Market Equilibrium}).

We evaluated the ``vanilla'' policymaker ($w_{h} = w_{b} = w_{s} = w_{f} = 1$), and four extreme cases where we optimize only one objective, i.e., (i) $w_{h} = 1$, (ii) $w_{b} = 1$, (iii) $w_{s} = 1$, and (iv) $w_{f} = 1$ (the rest of the weights are set to 0). The latter offers clear-cut results, but -- as we will show in Section \ref{Results Harvester Revenue and Social Welfare} -- it can potentially lead to adverse effects. In practice, the use of simulations can enable the testing of economic policies at \emph{large-scale}, and the ability to evaluate a range of different parameters, allowing the \emph{designer to ultimately select the weights that optimize the desired combination of objectives}.

\subsubsection*{Statistical Significance}

All simulations were \emph{repeated $8$ times}. The graphs depict the average values over those 8 trials, and the shaded area represents one standard deviation of uncertainty. The reported numerical results in the Tables are the average values of the last 400 episodes over those trials. (MA)DRL also lacks common practices for statistical testing \cite{AAAI1816669,hernandez2019survey}. In this work, we opted to use the Student's T-test \cite{student1908probable} due to its robustness \cite{colas2019hitchhiker}; p-values can be found in Appendix \ref{supplement: Additional Simulation Results}. All of the reported results that improve the baseline have p-values $< 0.05$.

\subsubsection*{Reproducibility}

Reproducibility is a major challenge in (MA)DRL due to different sources of stochasticity. To minimize those sources we used RLlib, an open-source MADRL library. See Appendix \ref{supplement: Agent Architecture} for details. The \emph{source code} can be found at \url{https://github.com/panayiotisd/fisher-market}.

\subsection{Comparing the ``Vanilla'' Policymaker to the Market Equilibrium Prices (MEP)} \label{Results Comparing the `Vanilla' Policymaker to the Market Equilibrium Prices}

Figures \ref{fig: harvester_reward} and \ref{fig: buyers_utility} depict the per-harvester mean reward and per-buyer mean utility, respectively, while rows 1 and 2 of Table \ref{tb: numerical results} show the relative difference of the achieved social welfare (sum of utilities), as compared to the market equilibrium prices (MEP). The vanilla policymaker (blue line in the figures and first column of the table) achieves results comparable to the MEP prices in both cases, with a loss of only $\approx7\%$ of social welfare. Similar results are achieved in the case of fairness -- both for the sellers and buyers (last two rows of Table \ref{tb: numerical results}) -- with both the MEP and the policymaker achieving a fair allocation (Jain index $\geq 0.98$\footnote{\label{footnote fairness}The higher the better; Jain index of 1 indicates a fair allocation.}). This is particularly important, as the market equilibrium is geared by design to optimize the aforementioned metrics, i.e., fairness for the participants and economic efficiency. Notably, the vanilla \emph{policymaker significantly outperforms the MEP when it comes to sustainability}, as we describe in Section \ref{Results Sustainability}.

\begin{tableorg}[t!]
\centering
\caption{Numerical results of the last 400 episodes of each training trial (averaged over the 8 trials). Each column represents the relative difference ($\%$) of the particular configuration of the policymaker, as compared to the market equilibrium prices ($100 (X_{\text{policymaker}} - Y_{\text{market eq.}}) /  Y_{\text{market eq.}}$), for each of the metrics presented in each row. The first column refers to the vanilla policymaker, where each objective in the reward function has the same weight (see Section \ref{Results}), and each of the following 4 columns refers to a policymaker that only optimizes the specific objective in the title (having weight 0 for the rest). Finally, the last three columns refer to a vanilla policymaker with obfuscated valuations (valuations split into 10 bins), noisy efforts ($10$\% uniform noise), and noisy efforts and valuations ($10$\% uniform noise), see Section \ref{Valuation Obfuscation}.}
\resizebox{1\linewidth}{!}{%
\begin{tabular}{@{}rrrrrrrrr@{}}
\toprule
\multicolumn{1}{l}{}       & \multicolumn{8}{c}{Policymaker}                                                           \\
                           & Vanilla & $w_{h} = 1$ & $w_{b} = 1$ & $w_{s} = 1$ & $w_{f} = 1$ & $\approx v$ & $\approx \phi$ & $\approx v$ \& $\phi$ \\ \midrule
Harvesters' SW                                        & -7.44                       & -1.74                           & -72.91                          & -31.37                          & -34.14                          & -9.71                          & -7.08                          & -9.24  \\
Buyers' SW                                            & -7.01                       & -24.71                          & 15.42                           & 1.23                            & 2.88                            & -11.51                         & -5.19                          & -10    \\
Stock Difference\footnoteref{footnote sustainability} & -15.3                       & -2.64                           & -10.58                          & -21.83                          & -12.99                          & -21.73                         & -19.59                         & -24.58 \\
Harvesters' Fairness                                  & -0.61                       & -0.05                           & -0.64                           & -0.72                           & -0.14                           & -1.04                          & -0.53                          & -0.77  \\
Buyers' Fairness                                      & -0.12                       & -0.18                           & -0.05                           & -0.09                           & -0.07                           & -0.31                          & -0.09                          & -0.31  \\ \bottomrule
\end{tabular}%
}
\label{tb: numerical results}
\end{tableorg}

\subsection{Harvesters' and Buyers' Social Welfare (SW)} \label{Results Harvester Revenue and Social Welfare}

Optimizing specifically for the harvesters' revenue or the buyers' utility (setting $w_{h} = 1$ or $w_{b} = 1$, respectively, and the remaining weights to 0), results in the policymaker closing the gap, or even significantly outperforming the MEP (green line in Figures \ref{fig: harvester_reward} and \ref{fig: buyers_utility} and second and third column of Table \ref{tb: numerical results}). The harvesters' Social Welfare (SW) improves from $-7.44\%$ to $-1.74\%$, while the buyers' SW exhibits a dramatic improvement from $-7\%$ to $+15.42\%$.

It is important to note, though, that contrary to the case of optimizing the sustainability or the fairness, exclusively optimizing the harvesters' SW has detrimental effects on the buyers' SW and vice versa (see Table \ref{tb: numerical results}). This is because these two objectives are somewhat orthogonal; low prices lead to high buyers SW but low harvesters SW, and vice versa (although money does not have an intrinsic value in Fisher markets). In this work, we showcase the potential of a vanilla policymaker, and the extreme cases of optimizing just one objective; it is up to the designer to ultimately select the weights that best serve the desired combination of objectives.

\subsection{Robustness} \label{Results Robustness}

Buyers' valuations are hard to elicit, while harvesters might experience perturbations to the actual effort exerted (see Section \ref{Robustness}). To evaluate this, we report results on noisy buyers' valuations (see Section \ref{Valuation Obfuscation}), split into 50 and 10 bins (third from the last column of Table \ref{tb: numerical results}, and Figure \ref{fig: buyers_utility_noisy_valuations_bins}; see Appendix \ref{supplement: Additional Simulation Results} for the rest). Noisy valuations lead to only a small drop in the buyers' and harvesters' SW ($\approx 2 - 4 \%$), the fairness remains the same, while sustainability improves significantly (up to $8\%$ compared to the vanilla policymaker). This comes to show that the policymaker is \emph{robust to noisy valuations}, which are much easier and more practical to elicit. Similar results were achieved for noisy efforts, and both noisy efforts and noisy valuations (last two columns of Table \ref{tb: numerical results}, and Appendix \ref{supplement: Additional Simulation Results}).

\subsection{Fairness} \label{Results Fairness}

The MEP are geared towards optimizing fairness; it is important to ensure that the introduction of the policymaker does not result in an exploiter-exploitee situation. All of the versions of the policymaker achieve a fair allocation (Jain index $\geq 0.98$,\footnoteref{footnote fairness} see Appendix \ref{supplement: Additional Simulation Results} for the other metrics). The relative values (Table \ref{tb: numerical results}) show a consistent improvement when specifically optimizing for fairness ($w_{f} = 1$) but, in absolute terms, all versions actually result in fair allocations.

\subsection{Sustainability} \label{Results Sustainability}

We measure sustainability as the maximum negative deviation from the equilibrium stock (see Section \ref{Multi-objective Optimization}). The introduced policymaker results in the emergence of \emph{significantly and consistently more sustainable harvesting strategies}. Figure \ref{fig: stock_difference} shows that the MEP maintain a stock that is $34\%$ below the equilibrium population (on average), while the policymaker is only $28.5\%$. Optimizing for sustainability ($w_{s} = 1$, green line) improves the difference to $26\%$.

More interesting is Figure \ref{fig: stock_difference_scarce}, where we simulate a \emph{scarce resource} environment (see Section \ref{Common Fishery Settings}). The introduction of the policymaker results in a \emph{dramatic improvement in sustainability}. The MEP maintain a population stock that is $97.3\%$ below the equilibrium population (on average), while the policymaker is $82.1\%$ and optimizing for sustainability improves the difference to $63.3\%$; almost $35\%$ improvement compared to MEP. In this setting, the MEP fail to result in a sustainable strategy and permanently \emph{deplete} the resources in $9.79\%$ of the episodes, with episodes lasting as low as $48$ time-steps (out of 500, which is the maximum possible). In contrast, the vanilla policymaker fails in $4.59\%$ of the episodes (min episode length of 180 time-steps), and the version that optimizes sustainability fails in only $2.24\%$ of the episodes (min episode length of 258 time-steps).

Importantly, optimizing for sustainability does not have detrimental effects on most other objectives, as seen in Table \ref{tb: numerical results}. The harvesters' and buyers' fairness improve as well, and so does the buyers' welfare. Only the harvesters' welfare degrades; but, as mentioned, it is up to the designer to select weights that best serve the desired combination of objectives.\footnote{\label{footnote sustainability}Note that the stock difference has negative values (negative deviation from the equilibrium stock) thus, in this metric, large negative numbers denote better performance of the policymaker.}

\subsection{Wasted Resources and Leftover Budget} \label{Results Wasted Resources Leftover Budget}

Figure \ref{fig: wasted} shows the percentage of wasted resources (harvested resources that remain unsold). Of course, by design, the MEP sell the entire harvest. Optimizing for sustainability results in a decrease of the wasted resources from $14\%$ to $10\%$ (blue vs. green line).

Regarding the buyers' leftover budget (see Appendix \ref{supplement: Additional Simulation Results}), the vanilla policymaker leaves $6\%$ of the budget unused, while optimizing for the harvesters' revenue leaves only $0.6\%$ of the budget (by design, the MEP use the entire budget).

The wasted resources and the leftover budget represent, in a sense, the over-supply and over-demand of the policymaker's allocation. It is clear by the low values for both metrics, and the results reported so far, that our policymaker reaches allocations that are \emph{qualitatively close to the market equilibrium} (at least in regard to the reported metrics), while \emph{optimizing for negative externalities} (such as sustainability).

\subsection{Extent of Intervention} \label{Results Extent of Intervention}

Finally, we want to concretely quantify the level of intervention needed to carry out the computed prices. The average (across the 500 time-steps of the final episode) relative price difference ($p^{p}_{r,t} - p^{ME}_{r,t} / p^{ME}_{r,t}$, where $p^{p}_{r,t}$ ($p^{ME}_{r,t}$) denotes the policymaker's (ME) price for resource $r$ at time-step $t$) for the vanilla policymaker ranges from $170\% - 437\%$ (for the 4 resources). We can further decrease this difference, by optimizing an additional objective, i.e., adding the following term to Equation~\ref{Eq: policymaker's reward}: $ - w_{i} |p^{p}_{r,t} - p^{ME}_{r,t}|$, where $w_{i}$ is a hyperparameter.\footnote{Of course, one can use a more involved function of the prices.} Adding this to the vanilla policymaker (all weights 1), reduces the relative price difference to $17\% - 20\%$. We also discretized the price difference into three bins: low ($\leq5\%$), medium ($5\% - 20\%$), and high ($>20\%$) intervention. The vast majority of the instances ($\approx 65\%$) are in the first two bins (the split amongst the bins is $(18.99\%, 45.42\%, 35.59\%)$).

In most practical applications, the ME prices required for the aforementioned optimization can not be known in advance. What is known, though, is the current market price of a resource. Thus, we can use the latter to ensure that the prices produced by the policymaker will only require a small intervention to the state of the market.

\section{Conclusion} \label{Conclusion}

We proposed a practical approach to computing market prices and allocations via \emph{deep reinforcement learning}, allowing us to counteract negative environmental externalities. We demonstrate significant improvements, especially towards solving the challenge of \emph{sustainability} of common-pool resources. Our work constitutes an important first step in studying markets composed of \emph{learning} agents, which are becoming ubiquitous in recent years.

\clearpage


\begin{appendices}

\section{Agents' Architecture: Additional Details} \label{supplement: Agent Architecture}

\subsection{Implementation, Hyperparameters, and Reproducibility} \label{supplement: Implementation, Hyperparameters, and Reproducibility} 

Reproducibility is a major challenge in (MA)DRL due to different sources of stochasticity, e.g., hyperparameters, model architecture, implementation details, etc. \cite{AAAI1816669,hernandez2019survey,Engstrom2020Implementation}. Recent work has demonstrated that code-level optimizations play an important role in performance, both in terms of achieved reward and underlying algorithmic behavior \cite{Engstrom2020Implementation}. To minimize those sources of stochasticity -- and given that the focus of this work is in the performance of the introduced policymaker and not of the training algorithm -- we opted to use RLlib\footnote{RLlib (\url{https://docs.ray.io/en/latest/rllib.html}) is an open-source library on top of Ray (\url{https://docs.ray.io/en/latest/index.html}) for Multi-Agent Deep Reinforcement Learning \cite{Liang2017Ray}.} as our implementation framework. All the hyperparameters were left to the default values specified in Ray and RLlib\footnote{See \url{https://docs.ray.io/en/latest/rllib-algorithms.html\#ppo}.}. For completeness, Table \ref{tab: List of hyperparameters} presents a list of the most relevant of them.

\begin{table}[!ht]
\centering
\caption{List of hyperparameters.}
\label{tab: List of hyperparameters}
\begin{tabular}{@{}lc@{}}
\toprule
\textbf{Parameter}                & \textbf{Value} \\ \midrule
Learning Rate ($\alpha$)          & 0.0001         \\
Clipping Parameter                & 0.3            \\
Value Function Clipping Parameter & 10.0           \\
KL Target                         & 0.01           \\
Discount Factor ($\gamma$)        & 0.99           \\
GAE Parameter Lambda              & 1.0            \\
Value Function Loss Coefficient   & 1.0            \\
Entropy Coefficient               & 0.0            \\ \hline   
\end{tabular}%
\end{table}

\subsection{Termination Condition} \label{supplement: Termination Condition}

An episode terminates when either (a) a fixed number of time-steps $T_{\max} = 500$ is reached, or (b) any of the resources gets depleted, i.e., the stock falls below a threshold $\delta = 10^{-4}$. We trained our agents for $2400$ episodes.

\subsection{Computational Resources} \label{supplement: Computational Resources}

All the simulations were run on an Intel Xeon E5-2680 (Haswell) -- 12 cores, 24 threads, 2.5 GHz -- with 256 GB of RAM.

\section{Fairness Metrics} \label{supplement: Fairness Metrics}

We employed three well-established fairness metrics: the Jain index \cite{DBLP:journals/corr/cs-NI-9809099}, the Gini coefficient \cite{gini1912variabilita}, and the Atkinson index \cite{ATKINSON1970244}:

(a)  The Jain index \cite{DBLP:journals/corr/cs-NI-9809099}: Widely used in network engineering to determine whether users or applications receive a fair share of system resources. It exhibits a lot of desirable properties such as population size independence, continuity, scale and metric independence, and boundedness. For an allocation game of $N$ agents, such that the $n^{\text{th}}$ agent is allotted $x_n$, the Jain index is given by Equation \ref{eq: Jain Index}. $\mathds{J}(\mathbf{x}) \in [0, 1]$. An allocation $\mathbf{x} = (x_1, \dots, x_N) ^\top$ is considered fair, iff $\mathds{J}(\mathbf{x}) = 1$.
\begin{equation} \label{eq: Jain Index}
  \mathds{J}(\mathbf{x}) = \frac{\left(\underset{n = 1}{\overset{N}{\sum}} x_n\right) ^ 2}{N \underset{n = 1}{\overset{N}{\sum}}  x_n ^ 2}
\end{equation}

(b) The Gini coefficient \cite{gini1912variabilita}: One of the most commonly used measures of inequality by economists intended to represent the wealth distribution of a population of a nation. For an allocation game of $N$ agents, such that the $n^{\text{th}}$ agent is allotted $x_n$, the Gini coefficient is given by Equation \ref{eq: Gini coefficient}. $\mathds{G}(\mathbf{x}) \geq 0$. A Gini coefficient of zero expresses perfect equality, i.e., an allocation is fair iff $\mathds{G}(\mathbf{x}) = 0$.
\begin{equation} \label{eq: Gini coefficient}
  \mathds{G}(\mathbf{x}) = \frac{\underset{n = 1}{\overset{N}{\sum}} \underset{n' = 1}{\overset{N}{\sum}} \left| x_n - x_{n'} \right|}{2 N \underset{n = 1}{\overset{N}{\sum}}  x_n}
\end{equation}

(c) The Atkinson index \cite{ATKINSON1970244}: Is a measure of the amount of social utility to be gained by complete redistribution of a given income distribution, for a given $\epsilon$. In our work, we used $\epsilon = 1$. For an allocation game of $N$ agents, such that the $n^{\text{th}}$ agent is allotted $x_n$, the Atkinson index of $\epsilon = 1$ is given by Equation \ref{eq: Atkinson index}. $\mathds{A}(\mathbf{x}) \in [0, 1]$. An Atkinson index of zero expresses perfect equality, i.e., an allocation is fair iff $\mathds{A}(\mathbf{x}) = 0$.
\begin{equation} \label{eq: Atkinson index}
  \mathds{A}(\mathbf{x}) = 1 - \frac{1}{\frac{\underset{n = 1}{\overset{N}{\sum}}  x_n}{N}} \left( \underset{n = 1}{\overset{N}{\prod}}  x_n  \right)^{1/N}
\end{equation}

\section{Societal Impact} \label{supplement: Societal Impact}

Sustainability and the preservation of the earth's natural resources constitute one of the most pressing issues and grand challenges in modern societies.\footnote{\label{fn: UN Goals}E.g., see \url{https://www.un.org/sustainabledevelopment/}.} For decades, the overarching priority of neoclassical economics has been economic growth \cite{barro2003economic}. Yet, the canonical assumptions of infinite and replenishable resources are practically unfounded and threaten the existence of the critical resources upon which our society ultimately depends. It is becoming increasingly clear that we need to shift our production patterns, decouple economic growth from environmental degradation, and increase resource efficiency.

Our approach can actively facilitate social mobility, sustainability, and fairness. As a potential negative social impact, the introduction of learning agents in socio-economic systems might bring forth an ``arms-race'' for the best means of production, which now shift from traditional, to computational resources and technological know-how. This can increase social inequality. Moreover, non-representative or biased modeling of markets and agents may lead to biased policy recommendations.

\section{Limitations} \label{supplement: Limitations}

The proposed policymaker is only partially interpretable, which can be a significant obstacle to real-world adoption of state-of-the-art ML-based decision systems. We therefore do not view our approach as immediately deployable. Instead, we see it as a practical tool for \emph{studying bounded interventions in complex markets}. In particular, the results of Section~\ref{Results Extent of Intervention} suggest that the learned policy can be implemented as a relatively small adjustment, or \emph{``nudge''}, to prevailing market prices, which may constitute a more realistic path toward adoption.

Another limitation is that the policymaker assumes access to buyer-side demand information, including an obfuscated representation of valuations. While this is standard in centralized market models and we partially relax it through market segmentation and noisy/coarsened valuation inputs (see Sections~\ref{Buyers Settings} and~\ref{Valuation Obfuscation}), it still assumes access to demand proxies that may not always be available in practice. We leave it open for future work to consider even weaker observation signals.

Additionally, as with all Deep Reinforcement Learning approaches, the proposed policymaker could be susceptible to adversarial attacks (e.g.,~\cite{Gleave2020Adversarial}). Furthermore, meaningful recommendations can only be acquired in environments where we can adequately model real-world production markets and their relevant attributes.

\begin{figure}[t!]
    \centering
    \includegraphics[width = 0.75 \linewidth, trim={0.7em 0.7em 0.7em 0.7em}, clip]{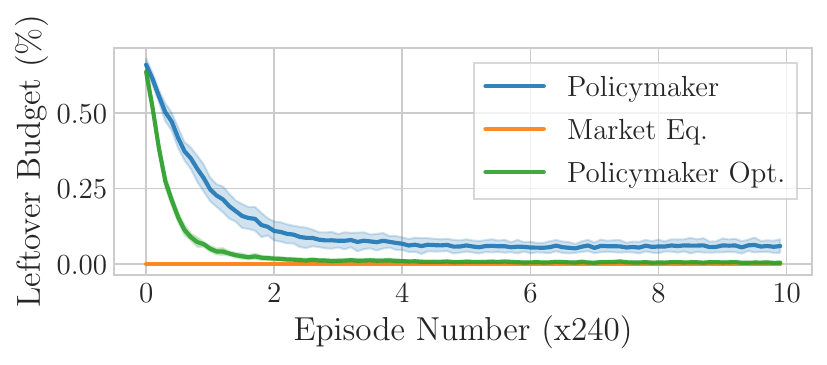}
    \caption{Evolution of the mean leftover budget over the number of training episodes. The orange line is the baseline (market equilibrium prices). The blue line refers to the vanilla policymaker where each objective in the reward function has the same weight. The green line refers to the policymaker that optimizes for the harvesters' reward (i.e., $w_{h} = 1$ and the rest of the weights to 0). Shaded areas represent one standard deviation.}
    \label{fig: Leftover Budget}
\end{figure}

\begin{table}[t!]
\centering
\caption{Numerical results of the last 400 episodes of each training trial (averaged over the 8 trials). Each odd column represents the relative difference ($\%$) of the vanilla policymaker, as compared to the market equilibrium prices ($100 (X_{\text{policymaker}} - Y_{\text{market eq.}}) /  Y_{\text{market eq.}}$), for each of the metrics presented in each row. Each even column shows the Student's T-test p-values. The first two columns present the results for an environment with 8 harvesters, 8 buyer classes, and 4 resources ($N = 8, R = 4, B = 8$), while the last two represent a larger scale scenario with 12 harvesters, 12 buyer classes, and 6 resources ($N = 12, R = 6, B = 12$). Note that the results of the first two columns are the same as the ones in Table \ref{tb: numerical results supplement}, and have only been included to facilitate comparison between the two test-cases.
}
\begin{tabular}{@{}lrrrr@{}}
\toprule
                              & \multicolumn{4}{c}{Vanilla Policymaker}                                                                          \\
                              & \multicolumn{1}{l}{$N = B = 8$} & \multicolumn{1}{l}{} & \multicolumn{1}{l}{$N = B = 12$} & \multicolumn{1}{l}{} \\
                              & $R = 4$                         & p-value              & $R = 6$                          & p-value              \\ \midrule
Harvesters' Social   Welfare  & -7.44                           & 1.21e-08             & -11.2                            & 4.15e-08             \\
Buyers' Social Welfare        & -7.01                           & 2.26e-06             & -4.51                            & 5.26e-05             \\
Stock Difference              & -15.30                          & 2.72e-09             & -19.31                           & 2.24e-16             \\
Harvesters' Fairness Jain     & -0.61                           & 3.71e-03             & -0.45                            & 6.37e-04             \\
Harvesters' Fairness Gini     & -2.78                           & 2.87e-04             & -2.4                             & 1.15e-05             \\
Harvesters' Fairness Atkinson & -0.29                           & 4.45e-03             & -0.21                            & 6.26e-04             \\
Buyers' Fairness Jain         & -0.12                           & 5.48e-05             & -0.04                            & 5.44e-06             \\
Buyers' Fairness Gini         & -1.49                           & 3.26e-07             & -0.78                            & 3.57e-08             \\
Buyers' Fairness Atkinson     & -0.06                           & 5.50e-05             & -0.02                            & 5.63e-06             \\ \bottomrule
\end{tabular}%
\label{tb: Large Scale Simulations}
\end{table}

\begin{table}[t!]
\centering
\caption{Numerical results of the last 400 episodes of each training trial (averaged over the 8 trials). The first column refers to the vanilla policymaker, while the second to the vanilla policymaker that also optimizes for the extent of the intervention to the market (see Section \ref{Results Extent of Intervention}).}
\begin{tabular}{@{}lrr@{}}
\toprule
                              & \multicolumn{2}{c}{Policymaker} \\
                              & Vanilla     & Interventions     \\ \midrule
Harvesters' Social   Welfare  & 64.38       & 65.59             \\
Buyers' Social Welfare        & 75.66       & 81.45             \\
Stock Difference              & -28.69      & -33.64            \\
Wasted Percentage             & 14.70       & 2.64              \\
Harvesters' Fairness Jain     & 0.98        & 0.99              \\
Harvesters' Fairness Gini     & 0.94        & 0.97              \\
Harvesters' Fairness Atkinson & 0.99        & 0.99              \\
Buyers' Fairness Jain         & 0.99        & 0.99              \\
Buyers' Fairness Gini         & 0.98        & 0.99              \\
Buyers' Fairness Atkinson     & 0.99        & 0.99              \\ \bottomrule
\end{tabular}%
\label{tb: interventions}
\end{table}

\section{Additional Simulation Results} \label{supplement: Additional Simulation Results} 

\paragraph{Results in Detail} In Table \ref{tb: numerical results supplement}, we provide a thorough account of the simulation results of the main text.

\paragraph{Leftover Budget} Figure \ref{fig: Leftover Budget} depicts the average (across the 8 trials) leftover budget over the number of training episodes.

\paragraph{Larger Scale Simulations} Table \ref{tb: Large Scale Simulations} presents the results of the larger simulation with 12 harvesters, 12 buyer classes, and 6 resources ($N = 12, R = 6, B = 12$). Note that the policymaker achieves similar results as in the smaller test-case (first two columns of Table \ref{tb: Large Scale Simulations}).

\paragraph{Extent of Intervention} In Table \ref{tb: interventions} we show that optimizing for the extent of the intervention to the market does not significantly affect the results for the original objectives.

\begin{landscape}
\begin{tableorg}[t!]
\centering
\caption{Numerical results of the last 400 episodes of each training trial (averaged over the 8 trials).\\Each odd column represents the relative difference ($\%$) of the particular configuration of the policymaker, as compared to the market equilibrium prices ($100 (X_{\text{policymaker}} - Y_{\text{market eq.}}) /  Y_{\text{market eq.}}$), for each of the metrics presented in each row.\\Each even column shows the Student's T-test p-values.\\The first two columns refer to the vanilla policymaker, where each objective in the reward function has the same weight (see Section \ref{Results}), and each of the following 8 columns refer to a policymaker that only optimizes the specific objective in the title (having weight 0 for the rest).\\Finally, the last 16 columns refer to a vanilla policymaker with obfuscated valuations (see Section \ref{Valuation Obfuscation}). The first 4 of them split the valuations into 50 and 10 bins, respectively, while the last 4 add uniform noise ($5\%$ and $10\%$, respectively).\\The p-values measure the significance of the difference of the policymaker results compared to the MEP.\\Finally, note that the stock difference has negative values (negative deviation from the equilibrium stock) thus, in this metric, large negative numbers are \emph{in favor} of the policymaker.}
\resizebox{0.7\paperheight}{!}{%
\begin{tabular}{@{}rrrrrrrrrrrrrrrrrrrrrrrrrrrr@{}}
\toprule
\multicolumn{1}{l}{}          & \multicolumn{26}{c}{Policymaker} \\
                              & vanilla & p-value & $w_{h} = 1$ & p-value & $w_{b} = 1$ & p-value & $w_{s} = 1$ & p-value & $w_{f} = 1$ & p-value & $\approx v$ (50) & p-value & $\approx v$ (10) & p-value & $\approx v$ (0.05) & p-value & $\approx v$ (0.1) & p-value & $\approx \varepsilon$ (0.05) & p-value & $\approx \varepsilon$ (0.1) & p-value & $\approx v$ \& $\varepsilon$ (0.05) & p-value & $\approx v$ \& $\varepsilon$ (0.1) & p-value \\ \midrule
Harvesters' Social   Welfare  & -7.44                       & 1.21E-08                    & -1.74                           & 1.37E-08                    & -72.91                          & 1.95E-18                    & -31.37                          & 2.47E-07                    & -34.14                          & 7.02E-08                    & -11.35                         & 3.81E-23                    & -9.71                          & 1.17E-22                    & -8.2                           & 1.85E-21                    & -10.07                        & 1.38E-21                    & -8.14                          & 1.56E-13                    & -7.08                         & 9.22E-10                    & -8.95                          & 1.40E-13                    & -9.24                         & 8.85E-15                    \\
Buyers' Social Welfare        & -7.01                       & 2.26E-06                    & -24.71                          & 1.03E-10                    & 15.42                           & 4.94E-19                    & 1.23                            & 3.95E-01                    & 2.88                            & 4.27E-02                    & -9.73                          & 4.90E-23                    & -11.51                         & 1.16E-24                    & -13.68                         & 4.54E-26                    & -11.56                        & 4.67E-25                    & -5.98                          & 2.23E-12                    & -5.19                         & 9.48E-13                    & -11.89                         & 5.43E-10                    & -10                           & 1.95E-13                    \\
Stock Difference              & -15.3                       & 2.72E-09                    & -2.64                           & 5.53E-04                    & -10.58                          & 1.00E-05                    & -21.83                          & 5.48E-10                    & -12.99                          & 2.59E-08                    & -23.4                          & 1.42E-17                    & -21.73                         & 4.59E-17                    & -24.68                         & 6.90E-18                    & -22.28                        & 2.79E-17                    & -16.51                         & 5.91E-15                    & -19.59                        & 1.51E-14                    & -23.27                         & 2.28E-13                    & -24.58                        & 3.41E-13                    \\
Harvesters' Fairness Jain     & -0.61                       & 3.71E-03                    & -0.05                           & 6.84E-03                    & -0.64                           & 6.22E-04                    & -0.72                           & 3.83E-03                    & -0.14                           & 2.95E-03                    & -1.16                          & 2.11E-20                    & -1.04                          & 8.01E-19                    & -1.33                          & 3.13E-20                    & -1.76                         & 2.06E-21                    & -0.56                          & 7.56E-05                    & -0.53                         & 1.50E-05                    & -0.8                           & 1.12E-02                    & -0.77                         & 7.70E-05                    \\
Harvesters' Fairness Gini     & -2.78                       & 2.87E-04                    & -0.54                           & 1.66E-02                    & -2.86                           & 1.37E-05                    & -3.08                           & 2.42E-05                    & -1.29                           & 1.61E-05                    & -4.57                          & 4.63E-16                    & -4.09                          & 2.46E-15                    & -4.75                          & 2.99E-16                    & -5.52                         & 3.86E-17                    & -2.78                          & 7.96E-06                    & -2.82                         & 7.18E-07                    & -3.27                          & 2.88E-05                    & -3.27                         & 1.07E-05                    \\
Harvesters' Fairness Atkinson & -0.29                       & 4.45E-03                    & -0.03                           & 2.77E-03                    & -0.29                           & 6.17E-04                    & -0.33                           & 3.93E-03                    & -0.07                           & 5.91E-03                    & -0.59                          & 1.53E-20                    & -0.48                          & 7.65E-19                    & -0.66                          & 3.04E-20                    & -0.93                         & 2.48E-21                    & -0.28                          & 8.89E-05                    & -0.26                         & 1.56E-05                    & -0.38                          & 8.09E-03                    & -0.37                         & 9.96E-05                    \\
Buyers' Fairness Jain         & -0.12                       & 5.48E-05                    & -0.18                           & 1.84E-04                    & -0.05                           & 1.15E-03                    & -0.09                           & 2.67E-04                    & -0.07                           & 8.19E-05                    & -0.27                          & 3.08E-14                    & -0.31                          & 2.80E-13                    & -0.31                          & 1.18E-15                    & -0.31                         & 2.39E-13                    & -0.07                          & 3.12E-04                    & -0.09                         & 1.94E-05                    & -0.42                          & 1.60E-10                    & -0.31                         & 2.92E-04                    \\
Buyers' Fairness Gini         & -1.49                       & 3.26E-07                    & -1.96                           & 6.23E-07                    & -0.84                           & 1.53E-05                    & -1.29                           & 4.00E-06                    & -1.06                           & 1.06E-07                    & -2.57                          & 8.94E-17                    & -2.74                          & 3.05E-15                    & -2.81                          & 3.18E-18                    & -2.78                         & 1.07E-16                    & -1.15                          & 2.31E-06                    & -1.31                         & 2.90E-08                    & -3.19                          & 9.12E-13                    & -2.6                          & 2.55E-07                    \\
Buyers' Fairness Atkinson     & -0.06                       & 5.50E-05                    & -0.09                           & 1.89E-04                    & -0.02                           & 1.06E-03                    & -0.05                           & 2.85E-04                    & -0.03                           & 6.41E-05                    & -0.13                          & 2.86E-14                    & -0.15                          & 4.16E-13                    & -0.15                          & 1.12E-15                    & -0.15                         & 3.06E-13                    & -0.04                          & 3.21E-04                    & -0.04                         & 2.24E-05                    & -0.21                          & 4.09E-10                    & -0.15                         & 2.87E-04                    \\ \bottomrule
\end{tabular}%
}
\label{tb: numerical results supplement}
\end{tableorg}
\end{landscape}

\end{appendices}








\bibliography{jaamas2024}

\end{document}